\documentclass[structabstract]{aa}

\usepackage{xcolor}

\usepackage{savesym}
\usepackage{amsmath}
\savesymbol{iint}
\usepackage{txfonts}
\restoresymbol{TXF}{iint}
\usepackage{natbib}
\usepackage{graphicx}
\usepackage{rotating}
\usepackage[flushleft]{threeparttable}
\usepackage{float}
\usepackage{array}
\newcommand{\Msun}{$M_{\odot}$}

\newcommand{\Mstar}{$M_{\star}$\ }

\newcommand{\kms}{km s$^{-1}$}

\graphicspath{{./}}
\usepackage{graphicx,caption}
\usepackage{makecell}

\begin{document}
	
	\title{A MUSE Spectro-imaging Study of the Th~28 Jet: Precession in the Inner Jet\thanks{Based on Observations collected with MUSE and X-Shooter at the Very Large Telescope on Cerro Paranal (Chile), operated by the European Southern Observatory (ESO). Program IDs:60.A-9322(A) and 095.C-0175(A)}}

	\author{
		A. \,Murphy \inst{1}
		\and 
		C. \,Dougados \inst{2}
		\and 
		E. T. \,Whelan \inst{1}
		\and 
		F. \,Bacciotti \inst{3}
		\and 
		D. \,Coffey \inst{4}
		\and 
		F. \,Comer\'on \inst{5}
		\and 
		J. \,Eisl\"offel \inst{6}
		\and
		T. P. Ray \inst{7}}

	\institute{Maynooth University Department of Experimental Physics, National University of Ireland Maynooth, Maynooth, Co. Kildare, Ireland 
	\and  
	Univ. Grenoble Alpes, CNRS, IPAG, 38000 Grenoble, France
	\and
	INAF, Osservatorio Astrofisico di Arcetri, Largo Enrico Fermi 5, 50125 Firenze, Italy
	\and
	School of Physics, University College Dublin, Belfield, Ireland
	\and
	ESO, Karl-Schwarzschild-Strasse 2, 85748 Garching bei M\"unchen, Germany
	\and
	Th\"uringer Landessternwarte, Sternwarte 5, 07778 Tautenburg, Germany
	\and
	Dublin Institute for Advanced Studies, Ireland}
	
	\titlerunning{Spectro-imaging of the Th~28 Jet} 
	\date{}
	
	\abstract {Th~28 is a Classical T Tauri star in the Lupus 3 cloud which drives an extended bipolar jet. Previous studies of the inner jet identified signatures of rotation around the outflow axis, a key result for theories of jet launching. Thus this is an important source in which to investigate the poorly understood jet launching mechanism.}{In this study we investigate the morphology and kinematics of the Th~28 micro-jets, with the aim of characterizing their structure and outflow activity, using optical integral-field spectroscopy observations obtained with VLT/MUSE.}{We use spectro-imaging and position-velocity maps to investigate the kinematic and morphological features of the jet, and to obtain a catalogue of emission lines in which the jet is visible. A Lucy-Richardson deconvolution procedure is used to differentiate the structure of the inner micro-jet region in selected emission lines. Spatial profiles extracted perpendicular to the jet axis are fitted to investigate the jet width, opening angle and the evolution of the jet axis.}{We confirm the previously identified knot HHW$_{2}$ within the red-shifted jet and identify three additional knots in each lobe for the first time. We also find [O III]$\lambda$5007 emission from the blue-shifted micro-jet including the knot closest to the star. Proper motions for the innermost knots on each side are estimated to be 0\farcs35 yr$^{-1}$ and 0\farcs47 yr$^{-1}$ for the red- and blue-shifted jets respectively. Based on this we show that new knots are ejected on an approximate timescale of 10-15 years. Gaussian fitting to the jet axis centroids shows a point-symmetric wiggle within the inner portion of both micro-jets indicating precession of the jet. We use the jet shape to measure a precession period of 8 years, with a half-opening angle $\beta <$ 0.6$^{\circ}$.  This precession may provide an alternative explanation for the rotation signatures previously reported. }{ We find that these parameters are compatible with precession due to a brown dwarf companion orbiting at a separation of $\leq$ 0.3 au. Further observations with higher spatial resolution may help to clarify the source of this precession.}

	\maketitle 
	
	\section{Introduction}
	
	\begin{figure*}
		\centering
		\includegraphics[width=18cm, trim= 0cm 0cm 0cm 0cm, clip=true]{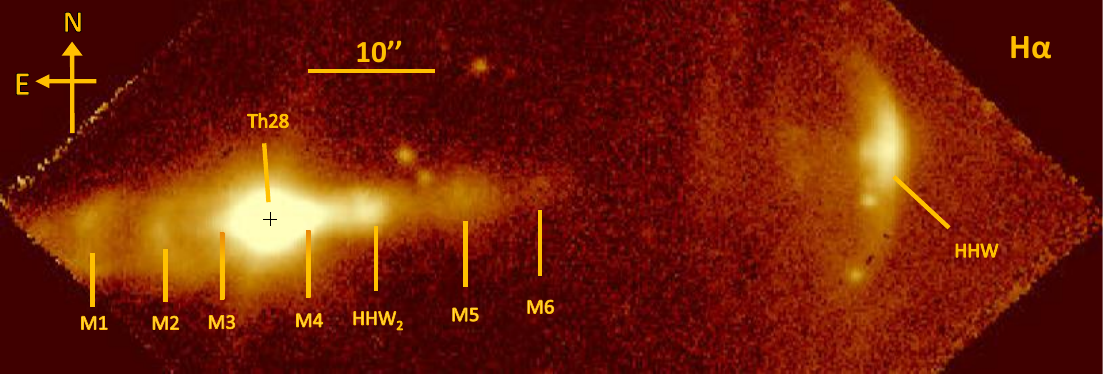}
		\caption{The MUSE observation of Th~28, showing the jet in the combined H$\alpha$ emission across all velocities before continuum subtraction. The positions of the star (black cross) and all knots discussed in this paper are indicated. A number of background stars are visible in the field, as well as the western Herbig Haro object associated with the red-shifted jet. Estimated FWHM of the PSF is 0\farcs9-1\arcsec.}
		\label{fig:th28snapshot}     
	\end{figure*}

	Jets are a key feature of young stellar objects (YSOs) for the majority of their pre-main sequence phase \citep{Whelan2014}. For example, Class 0, Class I and Class II low mass YSOs all drive jets \citep{Frank2014}. They are also seen across a range in masses from brown dwarfs to massive stars \citep{Whelan2012}. They are linked to accretion from the circumstellar disk \citep{Cabrit1990, Hartigan1995}, and are thought to play an important role in removing mass and angular momentum from the star-disk system as it forms \citep{Ray07}.  Observations of YSO jets on intermediate scales (100~au to 0.5~pc) expose the mass loss history of the jet and thus offer a window on the behaviour of the ejection/accretion system over a long time frame \citep{Frank2014}. Jet properties of interest are the spacing between the shocks in the jet (also referred to as Herbig-Haro (HH) objects and knots) and their proper motions, the evolution of the jet axis and the jet width with distance from the driving source. Variability in spacing and proper motion of the knots could point to variability in ejection speed and thus variable accretion activity \citep{Raga1990}. Periodic wandering in the jet axis or ``wiggling" can be caused by the presence of a close companion \citep{Anglada2007, Whelan2010}. By mapping the jet width the collimation of the jet can be studied, which is a key parameter for distinguishing between jet launching models \citep{Ray07}. Studies on these scales can also allow the connection between the different outflow components e.g. jets, wide-angled winds and cavities to be investigated \citep{Frank2014}.  
	
	Classical T Tauri stars (CTTS) are Class II low mass YSOs with an age of a few Myrs. They present ideal targets for  observations of jets on small and intermediate scales. They frequently possess jets traced by bright optical emission lines and are less densely embedded than younger protostars, making it feasible to observe the inner jet regions \citep{Dougados2000, Bacciotti2000, Whelan2004}. CTTSs are often thought of as only having micro-jets (jets extending to $<$ 1000 au from the source) due to the decline in outflow activity with age. However, studies have revealed that they can still drive large scale jets \citep{McGroarty2004}. 
	
	Spectro-imaging of CTTS jets in forbidden emission lines (FELs) has proved an important tool when it comes to advancing our knowledge of jet launching \citep{Lavalley2000, Maurri2014, Whelan2018}. Spectro-imaging allows for effective subtraction of the continuum emission, meaning the jet can be traced very close to the source. Furthermore, using this technique the morphology and physical conditions of the jets can be analysed in different velocity bins and past works have investigated jet collimation, wiggling, and velocity structure \citep{Agra2011, Maurri2014}. While the majority of spectro-imaging studies were done on small spatial scales (less than a few hundred au), using high resolution integral-field unit (IFU) instrumentation (e.g. GEMINI/NIFS and VLT/SINFONI), and mock-IFU observing techniques from space using HST spectroscopy \citep{Bacciotti2000, Bacciotti2002, Hartigan2004, Melnikov2009, Maurri2014}, this work can now be easily extended to intermediate scales, thanks to a recent addition to the European Southern Observatory's Very Large Telescope (ESO VLT) \citep{Schneider2020}. 
	
	The Multi-Unit Spectroscopic Explorer (MUSE) presents an exceptionally powerful tool for jet observations. This is a second-generation VLT instrument which provides integral-field spectroscopy observations in the 4000-9000~\AA\ range and over an exceptionally large 1\arcmin\ $\times$ 1\arcmin\ field of view (FOV) \citep{Bacon10}. It has 0\farcs2 pixel sampling and medium spectral resolution (R = 2000-4000). The key strength of this instrument lies in it offering simultaneous coverage of numerous optical emission lines, as well as continuum emission which enables the latter to be effectively estimated and subtracted. This powerful combination of features makes MUSE an ideal tool for studying jets on intermediate scales. In this paper we present MUSE observations of the bipolar jet of a well-known CTTS, Th~28, focusing on the morphology and kinematics of its micro-jets within 15\arcsec\ of the driving source. 
	
	Using channel maps and position-velocity (PV) diagrams we study the kinematics of both jet lobes, identify new knots and investigate the jet collimation and position of the jet axis with distance from the source. In Section 2 the target, observations and analysis of the datacube we undertook are described. In Section 3 the morphological and kinematical results are presented. In Section 4 we discuss our modelling of the jet axis trajectory. Our conclusions are summarised in Section 5.

	\begin{figure*}
		\centering
		\includegraphics[width=18cm, trim= 1cm 1cm 0cm 0cm, clip=true]{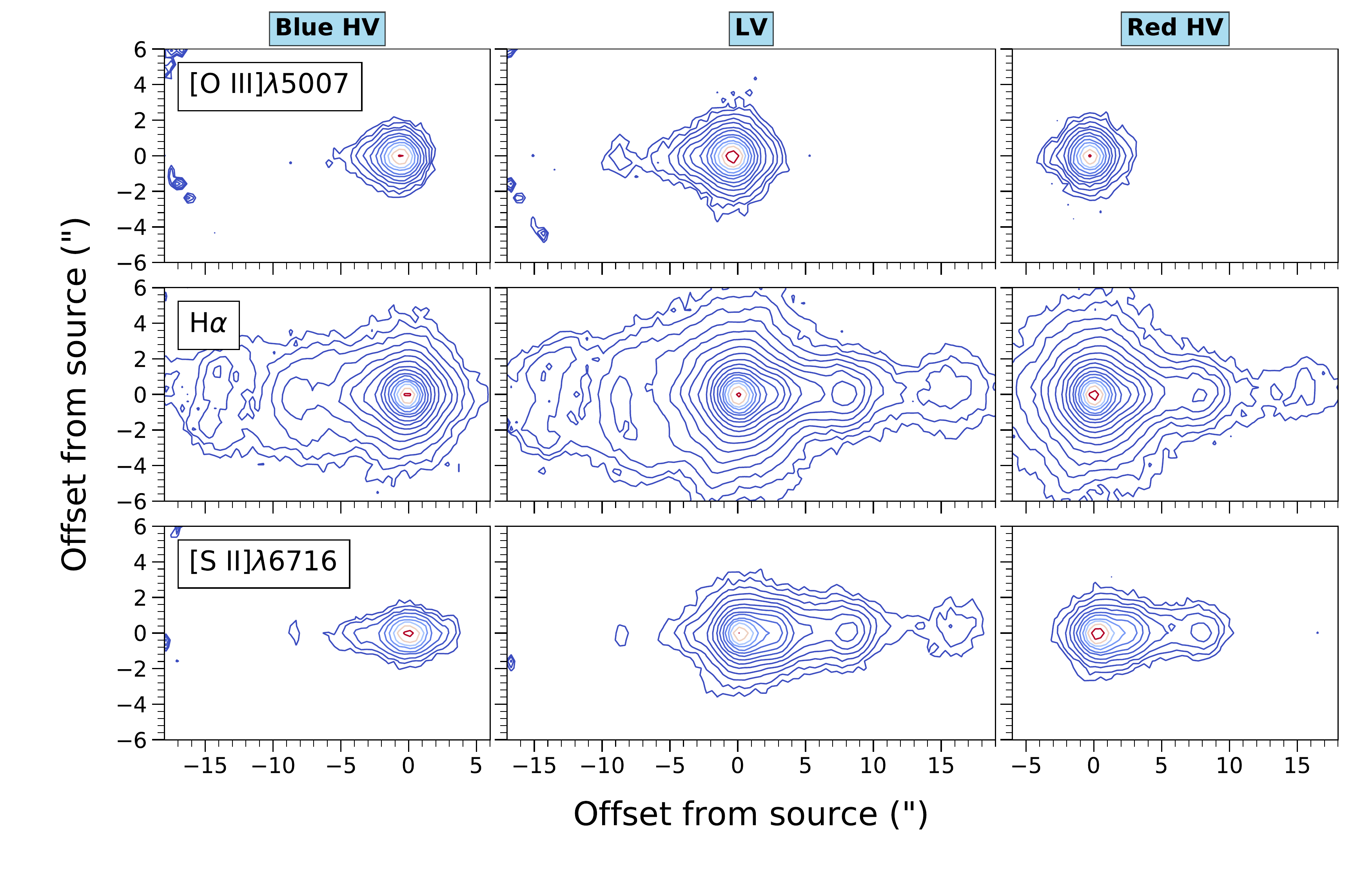}
		\caption{Spectro-images showing the Th~28 jet emission in selected emission lines, with contours starting at 3-$\sigma$ of the background noise (at an average flux of 16 $\times$ 10$^{-20}$ erg~s$^{-1}$~cm$^{-2}$ $\AA^{-1}$) and increasing as factors of $\sqrt{3}$. {\bf Left:} the high-velocity blue-shifted emission (-200 to -90 \kms). {\bf Centre:} the low-velocity emission (-90 to +90 \kms). {\bf Right:} the high-velocity red-shifted emission (+90 to +200 \kms).}
		\label{fig:main_specims}     
	\end{figure*}

	\section{Target, observations, and data}
	
	\subsection{Th~28}
	\label{section:th28_background}
	
	\object{ThA 15-28}  (hereafter Th~28; also referred to as Sz 102 or Krautter's Star) is a CTTS in the Lupus III cloud. It drives a bright bipolar jet (HH 228) in the plane of the sky \citep{Graham88}. We summarize the parameters of this target in Table \ref{table:target_params}.  Its stellar spectrum is heavily veiled with very weak absorption lines and may be seen primarily in scattered light \citep{Hughes1994}. The star is estimated to have a spectral type of approximately K2 and age $<$ 2 Myr. and is very under-luminous at 0.03 $L_{\odot}$ \citep{Mortier2011}. These properties pointed to it having an edge-on disk and recent ALMA observations do find the disk rotation axis to have a large inclination (\textit{i} $\geq$ 73$^{\circ}$) with respect to the line of sight \citep{Louvet2016}. 
	
	 Mass estimates for the star based on the H$\alpha$ accretion emission and ALMA observations of the disk vary between 1-2~$M_{\odot}$ \citep{Louvet2016, Comeron2010}. Estimated distances to Th~28 vary between 160~pc and 200~pc \citep{Wang09, Galli13}. Updated estimates of Th~28's neighbors and Lupus III as a whole based on GAIA parallax measurements indicate a distance of 162 $\pm$ 3 pc \citep{Dzib2018}. We therefore take the distance of Th~28 to be 160 pc.

\begin{table*}
	\centering
	\begin{threeparttable}
	\centering
	
	\caption[target]{\label{table:target_params} Table of key parameters for Th~28.}
	
	\begin{tabular}{{p{0.08\textwidth}<{\raggedright} p{0.07\textwidth}<{\raggedright} p{0.06\textwidth}<{\raggedright} p{0.05\textwidth}<{\raggedright} p{0.05\textwidth}<{\raggedright} p{0.1\textwidth}<{\raggedright} p{0.05\textwidth}<{\raggedright} p{0.06\textwidth}<{\raggedright} p{0.05\textwidth}<{\raggedright}}}
		\hline \hline
		Outflow & Location &  \makecell[l]{Distance \\ (pc )} & \makecell[l]{$v_{sys}$ \\ (\kms)} & \makecell[l]{Age \\(Myr)} & Spectral type & \makecell[l]{\Mstar~ \\(\Msun)} & \makecell[l]{$i_{disk}$ \\($^{\circ}$)} & \makecell[l]{Jet PA \\($^{\circ}$)} \\
		& & & & & & & & \\
		HH 228 & Lupus 3 & 160$^{2}$ & 2.9$^{3}$ & $<$ 2$^{4}$ & K2:$^{4}$ & 1.6$^{3}$ & 73$^{3}$ & 95$^{1}$ \\ 
		\hline
		
	\end{tabular} 
	
	\tablebib{(1) \cite{Graham88}; (2) \cite{Dzib2018}; (3) \cite{Louvet2016}; (4) \cite{Mortier2011}}
	\end{threeparttable}
	
\end{table*}

	The Th~28 jet lies in the east-west direction with position angle (PA) $=$ +95$^{\circ}$ for the blue-shifted lobe and -85$^{\circ}$ for the red-shifted lobe \citep{Graham88}. It is known to extend to a distance of at least 0.32~pc from the source (in the blue-shifted jet). The inclination of 73$^{\circ}$ estimated for the disk from the CO observations of \cite{Louvet2016} implies a jet axis inclination to the plane of the sky of $<$ 17$^{\circ}$ . It comprises the prominent red-shifted bowshock HH 228 W \citep{Graham88} as well as a knot within the micro-jet designated HHW$_{2}$ \citep{Comeron2010}, as well as a string of HH objects associated with the blue-shifted jet (e.g. HH 228 E$_{1}$ to HH 228 E$_{4}$). Note that from here on we drop the HH number 288 when referring to the different jet knots. The jet and counter jet are noticeably asymmetric. Radial velocities of the HH objects in the blue-shifted jet are greater by an average factor of 1.5-2 compared with those in the red-shifted jet \citep{Graham88, Comeron2010}. Proper motion studies of knots on both sides have also been carried out using H$\alpha$ and [S II] images \citep{Wang09, Comeron2011}. The blue-shifted knots have estimated proper motions of $~$0\farcs43~yr$^{-1}$ to 0\farcs5~yr$^{-1}$, whereas the red-shifted bowshock HHW shows a proper motion of 0\farcs4~yr$^{-1}$. This gives estimated jet velocities of 325~km s$^{-1}$ to 380~km s$^{-1}$ (\textit{D} = 160~pc) in the blue-shifted jet compared with 300~km s$^{-1}$ in the red-shifted jet. An interesting feature of this jet is revealed here. The difference in radial velocities between the red and blue lobes is significantly higher than the difference in tangential velocity; additionally, the tangential velocities would be considered high for HH objects \citep{Comeron2010}. 
	
	 Th~28 is notable for being one of a handful of jets in which transverse velocity gradients, a signature of jet rotation, have been detected using HST-STIS \citep{Coffey2004}. Consistent gradients were measured in optical lines observed with 0\farcs1 spatial resolution across the jet, although corresponding measurements in NUV lines with higher spatial resolution found inconclusive results. However, both lobes of the optical Th~28 jet are found to rotate in the opposite sense to the surrounding disk \citep{Louvet2016}, contrary to the launching models, raising questions about the possible cause.

	\subsection{Observations}
	The MUSE observations were made on 23rd June 2014, during the instrument's science verification phase. The average seeing was 0\farcs9 during the observations and the pixel scale was 0\farcs2. The observations were taken in Wide-Field Mode (WFM). The detector was rotated 45$^{\circ}$, aligning the jet axis with the diagonal of the detector to maximise the extent of the jet that was observed. As the WFM FOV is 1\arcmin $\times$ 1\arcmin, this allowed the observations to capture approximately 80\arcsec\ along the jet axis. The source was positioned so that HHW was included in the field and therefore more of the red-shifted than the blue-shifted jet was observed. The observations consisted of 6 exposures with a total exposure time of 3880s, and the wavelength range of the observations was 4570 \AA\ to 9350 \AA\ with a wavelength-dependent spectral resolution of between 80~km s$^{-1}$ (4750 \AA) and 40~km s$^{-1}$ (9350 \AA). The velocity resolution for key lines of interest in this paper range between 75~km s$^{-1}$ for the [O III]$\lambda$5007 line, to 56~km s$^{-1}$ for the [S II]$\lambda$ 6716 \AA\ and 6731 \AA\ lines. X-Shooter spectra taken on 25th June 2015 are also included in this work, for the purpose of proper motion estimates only. This X-Shooter data will be analysed in more detail in a separate paper.

	\subsection{Data reduction and analysis}
	\label{section:reduction}
	The observations were reduced using version 2.0.1 of the standard MUSE pipeline \citep{Weilbacher2020}; this removes the instrumental signature of each individual IFU and includes bias and dark subtraction, illumination corrections and wavelength calibration; the data from individual CCDs are then combined in post-processing (using the {\it muse$\_$scipost} recipe) after flux calibration, sky subtraction and astrometric calibration. The resulting exposures were then combined to produce a single datacube. Separate datacubes were produced for each 500~\AA\ band in the full wavelength range. 
	
	We have written a set of Python tools for the analysis of the MUSE datacubes. These tools enable the creation of datacube subsets and the easy selection of individual emission line regions, 2D slices (position-velocity and velocity channel maps) and 1D spectra. They also include routines for continuum subtraction and profile analysis (e.g. Gaussian fitting). Where Gaussian fits are used to obtain centroid positions in spatial position or velocity, the errors on the centroids are estimated as $\frac{0.4~\textup{FWHM}}{\textup{SNR}}$ of the fitted Gaussian \citep{Porter2004}. These tools were used to conduct the data analysis presented here.
	
	A sky spectrum summed across pixels in target-free regions of the FOV was created for each datacube subset; the catalogue of \cite{Hanuschik03} was used to identify residual telluric lines, which were then used to check the wavelength calibration in each region of the MUSE spectrum. Within the 6000~\AA\ - 7000~\AA\ band of interest in this study, the calibration was accurate to less then 15~$\%$ of the velocity resolution.
	
	To remove the continuum, we choose to rely on a simple subtraction of a baseline flux for two reasons. First, the emission closest to the star is dominated by scattered light. As a result, in attempting to sample the stellar spectrum we found strong contamination from scattered jet emission, making it very difficult to sample a clean stellar spectrum for subtraction. Second, the spectrum of Th~28 is heavily veiled  with little or no absorption visible. This results in an extremely flat continuum marked only by emission lines, making it highly suitable for a straightforward baseline subtraction. A routine was therefore written to sample the continuum on either side of a given emission line (covering a typical total range of 40-50 \AA) and hence to estimate a linear baseline for subtraction. The baseline subtraction procedure was used to produce a continuum-subtracted cube for each emission line of interest. The routine additionally stores a cube containing the estimated continuum, for inspection. 
	
	As the jet PA was at a slight angle to the horizontal axis of the datacubes, a rotation procedure was implemented. The three central wavelength pixels of the bright Ha emission were binned to create a combined image. We then used the bright, well-collimated red-shifted micro-jet as a guide to measure the PA: this portion of the image was binned in 2-pixel segments along the x-axis and the centroid at each position obtained by a Gaussian fit to the jet cross-section. The angle of the jet axis was then estimated by a linear fit through the centroids. The estimated angle of the axis to the horizontal was 5.2$^{\circ}$. For each data cube, a numpy interpolation routine was then used to rotate each wavelength slice through this angle and the slices recombined into a rotated cube. The fitting procedure was repeated on the H$\alpha$ cube to confirm that the rotated jet axis was close to horizontal. All the results in this study are obtained from these rotated cubes.
	
	\section{Results}
	
	Figure \ref{fig:th28snapshot} shows the MUSE view of the Th~28 jets and western bow shock in H$\alpha$ emission before continuum subtraction, including several background stars. In this paper, we focus on the Th~28 micro-jets within approximately $\pm$~15\arcsec\ of the source position: this region is referred to here as the inner jet. The large bow shock (HHW) located to the outer western edge of the observation will form the subject of a later paper. Analysis of the full MUSE cube reveals a wealth of FELs, as well as H$\alpha$ and H$\beta$, tracing the inner jets. See Table A.1 for a full list of these lines. Figure \ref{fig:main_specims} shows channel maps of the jet in three key lines, with channel maps for all the emission lines in which the jet is detected given in the appendix. In particular we note the detection of [O III]$\lambda$5007 emission in the blue-shifted jet. This line was previously reported by \citet{Comeron2010} and associated with the stellar wind; however, in this data the [O III] emission is clearly seen from the inner jet and the centre of the first bow shock M2.
	
	For the channel maps, emission was split into three velocity bins: a high-velocity blue-shifted (Blue HV) bin from -200~\kms\ to -90~\kms, a central low-velocity (LV) channel at +/- 90~km s$^{-1}$, and a high-velocity red-shifted (Red HV) channel from +90~\kms\ to +200~\kms. Due to the low inclination of the Th~28 jet of $<$ 17$^{\circ}$ the red- and blue-shifted jet lobes exhibit relatively small mean radial velocities, hence both jets can be observed in the central velocity channel. Emission from both can also be seen in the HV channels due to the FWHM of the line profiles (150 \kms\ and 260 \kms\ in the red- and blue-shifted jets respectively, measured in H$\alpha$ at +/- 2-3\arcsec~from the source). All velocities are quoted with respect to the systemic velocity of Th~28, measured by \cite{Louvet2016} to be +2.9~\kms. Over the wavelength range of interest, significant line emission is seen from around the star itself, which may be due to scattering of jet emission. As it was not possible to obtain an uncontaminated stellar spectrum, any photospheric contribution cannot be readily removed. However, many features of the inner jets can still be discerned.
	
	From Figure \ref{fig:main_specims}, the blue-shifted eastern jet appears much wider and more diffuse in H$\alpha$ emission, with two bow or bubble-shaped knot features located at -8\farcs8 and -14\farcs2 from the driving source (labelled here M1 and M2). These knots have not been described in previous observations of Th~28 and are discussed further in Section 3.3.1.  The FEL channel maps show the blue-shifted jet to be much more collimated in comparison to the wide H$\alpha$ flow. The inner 5\arcsec\ of this jet is also visible in highly red-shifted channels of H$\alpha$ and [O III]$\lambda$5007 emission with $v_{rad}$ $>$ +100~\kms, which may be due to scattering of red-shifted jet emission.  The bright red-shifted jet is detected in many more FELs than the blue-shifted jet (refer to Table A1 and atlas of channel maps), appearing as a collimated flow extending to a prominent knot at 7\farcs8 from the source (HHW$_{2}$). Visible in H$\alpha$, [N II] and [S II] are additional large knots at 16\arcsec\ and 21\farcs6 (labelled M5 and M6 here). These are discussed further below. 
	
	To further investigate the morphology and kinematic structures of the micro-jets, we use a deconvolution procedure to reduce the effect of seeing and improve the spatial resolution close to the source. This procedure and the results are described in the next subsection; we will then discuss the identification of the new knots within the micro-jets, the kinematic information revealed, and measurement of the width and centroid position along the axis of the micro-jets.

	\subsection{Deconvolution}
	\label{section:deconvolution}
	
		\begin{figure*}
		\centering
		\includegraphics[width=18cm, trim= 1cm 1cm 0cm 0cm, clip=true]{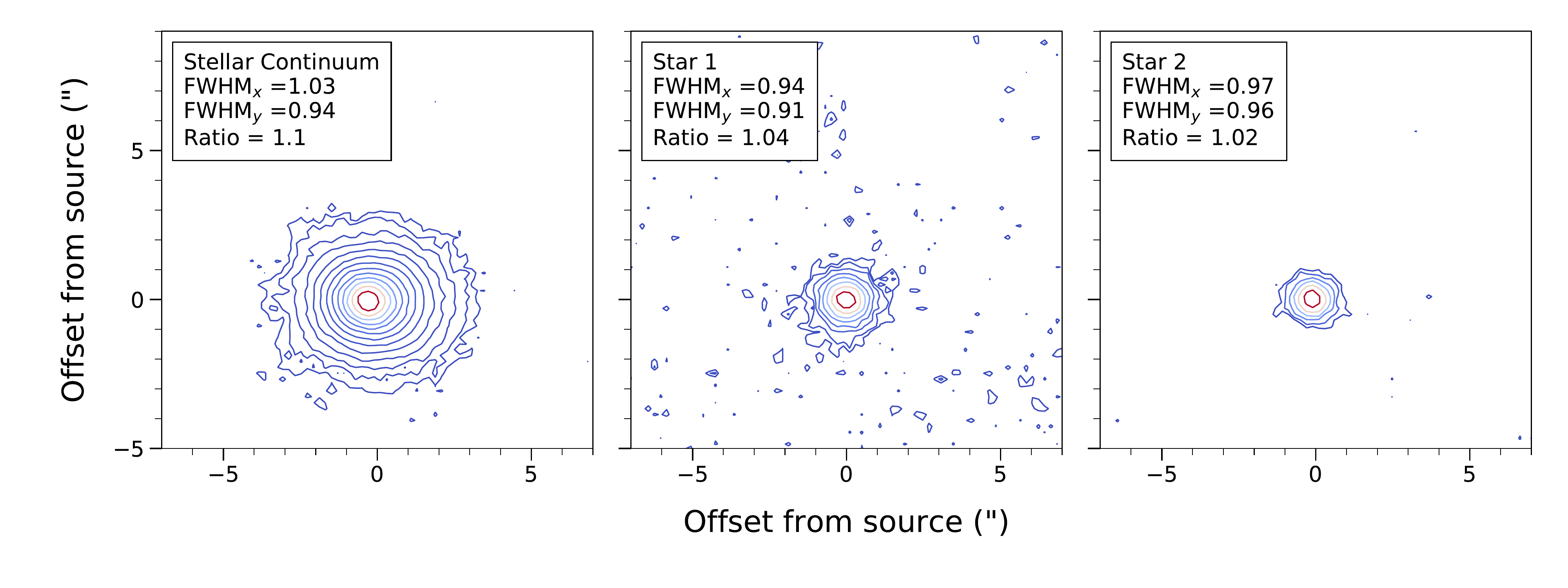}
		\caption{Spectro-images showing samples of the PSF estimated from continuum images of Th~28 (left) and two background stars (center, right) in the H$\alpha$ wavelength band, with contours starting at 4-$\sigma$ of the background noise and increasing as factors of $\sqrt{3}$. Measured FWHM values are given in the y and x directions (perpendicular to  and parallel with the jet direction, respectively).}
		\label{fig:psf_samples}     
	\end{figure*}
	
    To enhance detail in the jet regions closest to the star, a Richardson-Lucy deconvolution algorithm was applied to the continuum-subtracted data. Background regions of the continuum-subtracted image were sampled to estimate the residual background flux. This offset was subtracted from the image and included as a parameter in the deconvolution task.

    To obtain an estimate of the point-spread function (PSF), we sampled continuum images of Th~28 and the two brightest background stars in the FOV. In each case, the images were found as an average of the continuum emission on both sides of the emission line of interest. Figure \ref{fig:psf_samples} shows examples for the PSF in the H$\alpha$ region. To evaluate the PSF shape, we fitted each sample with a Gaussian function along cross-sections in the x and y directions (parallel to and across the jet axis direction). We find that the PSF consistently shows a slight elongation in the x direction; however, this is more pronounced in the continuum image of Th~28, reflecting a probable contribution from scattered light along the jet axis. 

    We therefore attempted to carry out deconvolution using as the PSF an image of a 2D Gaussian function with the FWHM values fitted from the brightest background star; this would avoid artefacts introduced by masking additional nearby background stars. However, the resulting deconvolved images suffered from heavy artefacting close to the star and the inner part of the micro-jets. We were instead able to avoid this by using the continuum image of Th~28 as the PSF. The average FWHM$_{y}$ was 0\farcs94, with average FWHM$_{x}$  1\farcs03 across and along the jet axis respectively, making the PSF slightly wider than the average estimated seeing of 0\farcs9. The impact of the PSF shape to the deconvolution results will be discussed in our conclusions.
    
    The spectro-image of the inner jet at each individual wavelength was then deconvolved with the estimated PSF for 20 iterations using the IRAF \textit{lucy} task, to produce a recompiled data cube of the deconvolved line emission at each wavelength. The average PSF FWHM estimated from the residual continuum emission at the source position was reduced from 1\farcs0 before deconvolution to 0\farcs44 after deconvolution. Figure A.1 presents the channel maps of the inner region of the jet before and after deconvolution.

	\subsubsection{New knot identification}
	\label{subsubsection:newknots}
	
\begin{figure}
		\includegraphics[width=9cm, trim= 0.4cm 2.3cm 0.4cm 0cm, clip=true]{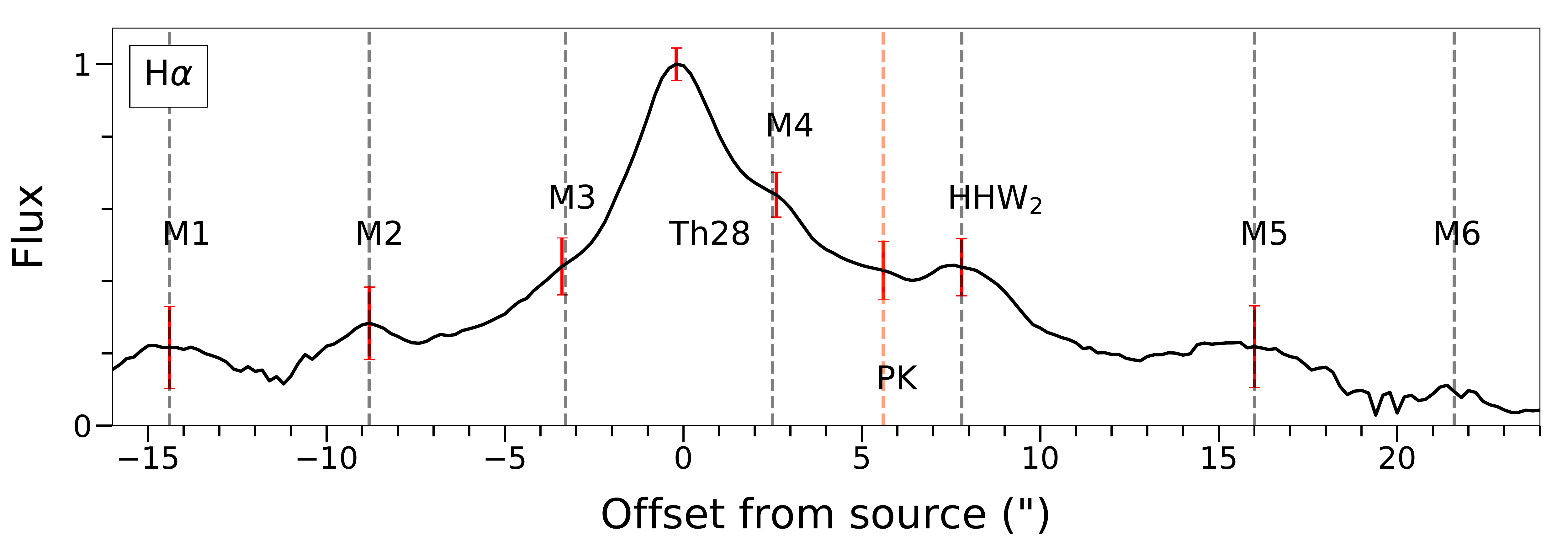}
		\includegraphics[width=9cm, trim= 0.4cm 2.3cm 0.4cm 0cm, clip=true]{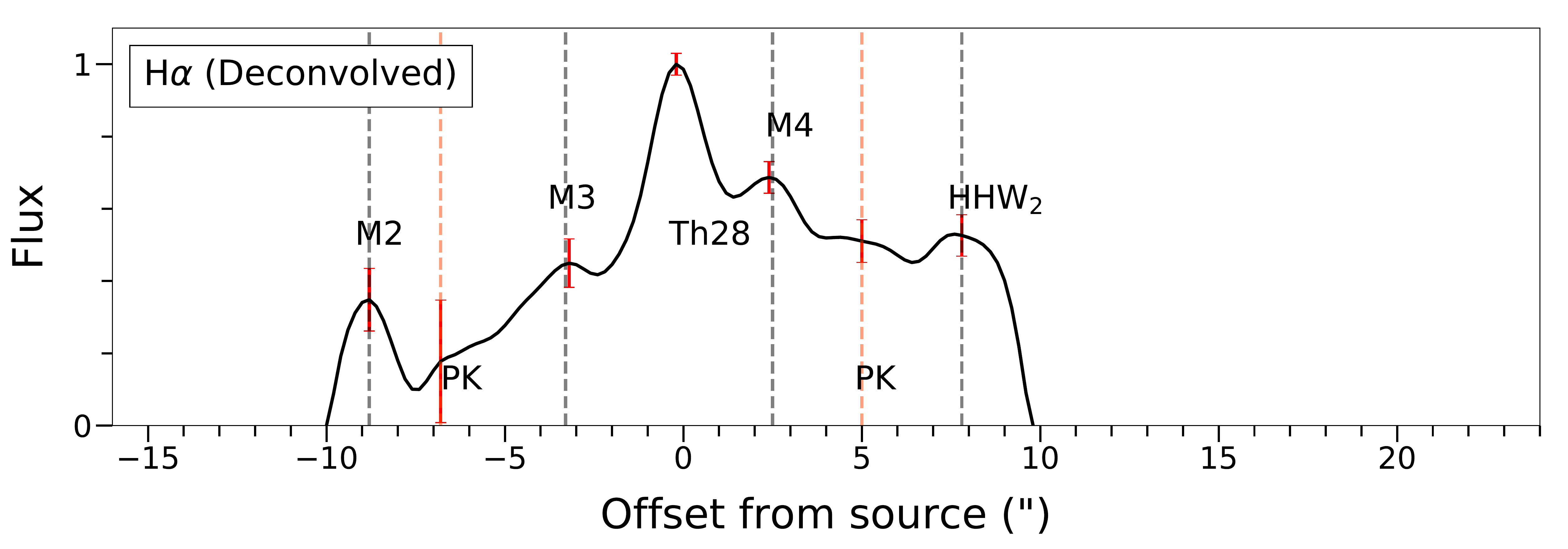}
		\includegraphics[width=9cm, trim= 0.4cm 2.3cm 0.4cm 0cm, clip=true]{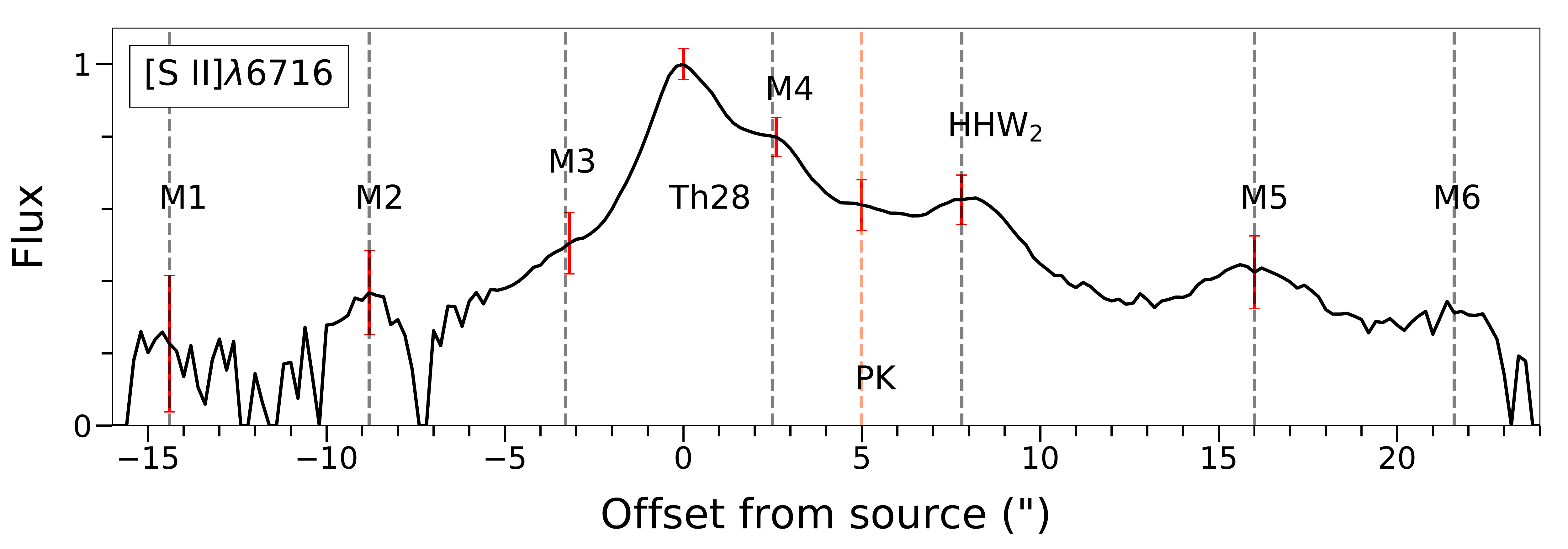}
		\includegraphics[width=9cm,  trim= 0.4cm 0cm 0.4cm 0cm, clip=true]{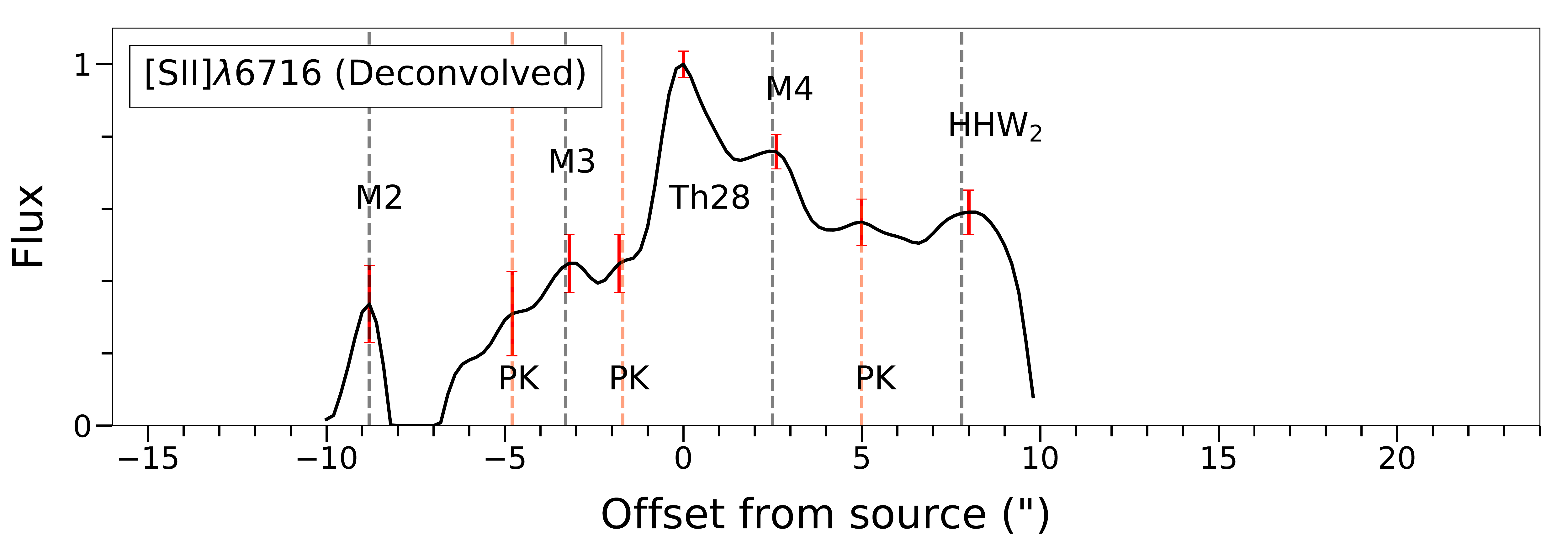}
		\caption{H$\alpha$ and [S II]$\lambda$6716 flux along the axis of the jet, taken from a horizontal slice through the jet centre after rotating the velocity-binned image by -5.2$^{\circ}$. Profiles are shown taken before and after deconvolution, with deconvolved profiles truncated at $\pm$ 10\arcsec. The flux axis is plotted on a log scale to prevent the plot being dominated by extremely high flux around the star and normalised to better compare the profiles in both lines. Possible knots are marked as PK.}
		\label{fig:axis_fluxprofile}     
	\end{figure}

	We report the detection of six new knots (labelled M1 to M6) not previously identified in the Th~28 jet. M1, M2, M5, and M6 are visible before deconvolution while M3, and M4 are only clearly detected after deconvolution. To further examine the new knots detected in the deconvolved image, we analyse profiles of the flux along the jet axis in H$\alpha$ and [S II] emission, both before and after deconvolution. Examples are shown in Figure \ref{fig:axis_fluxprofile}. In general, the peaks correspond well to the knots seen in the channel maps. Although the deconvolved profiles are limited to the inner 10\arcsec\ of the micro-jets, they clearly show knots within this region which were not obvious beforehand. These knots will now be discussed individually.

	\begin{figure}
		\includegraphics[width=9cm, trim= 2.2cm 1.5cm 3.6cm 3cm, clip=true]{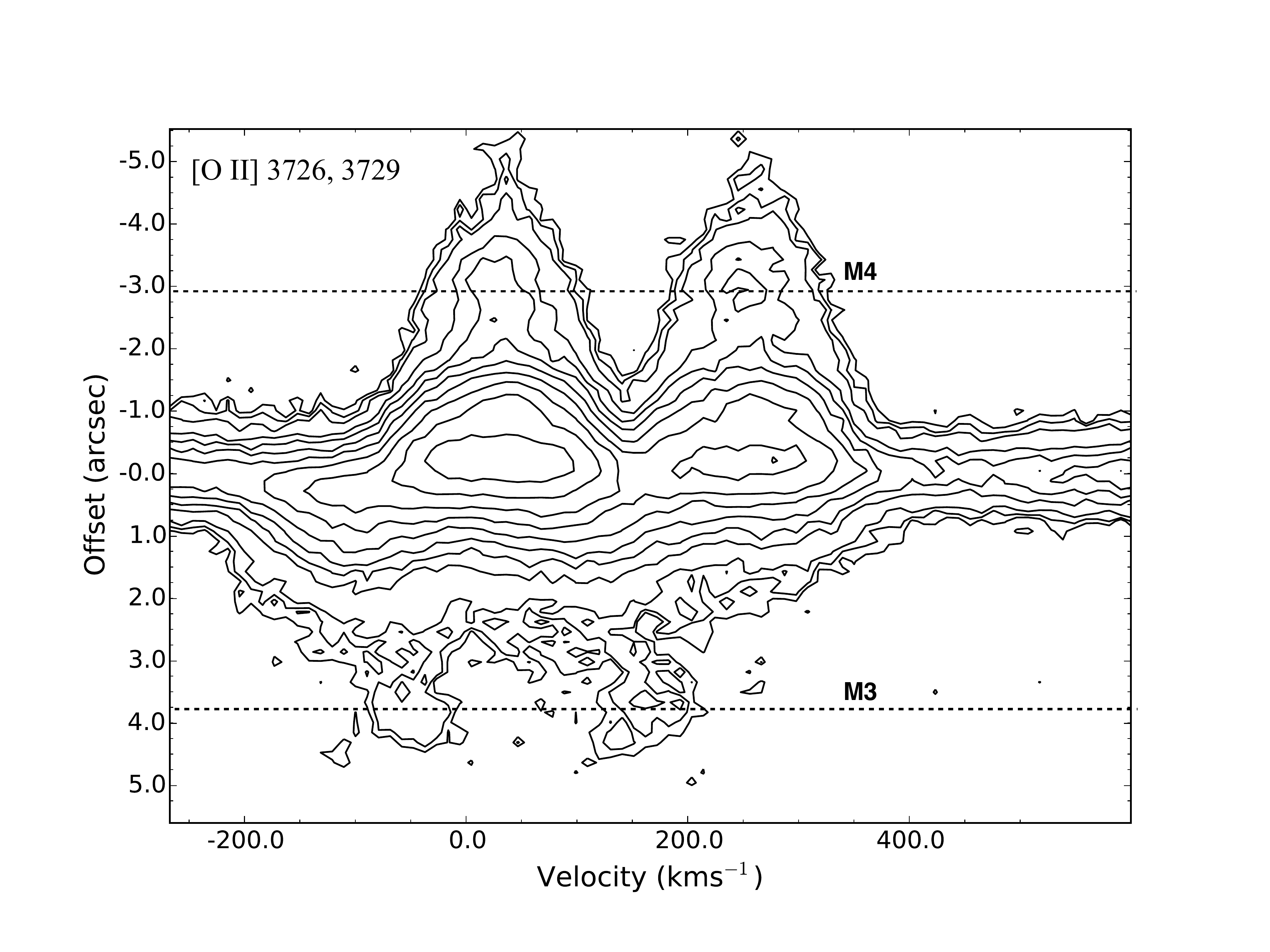}
		\caption{Position-velocity maps of the Th28 jet obtained with X-Shooter, showing the jet in [O II]$\lambda$3726, with [O II]$\lambda$3729 also on the right. Contours begin at 3-$\sigma$ and increase on a log scale.}
		\label{fig:xshooter_knots}     
	\end{figure}
	
	\begin{figure*}
		\includegraphics[width=18cm, trim= 0cm 0cm 0cm 0cm, clip=true]{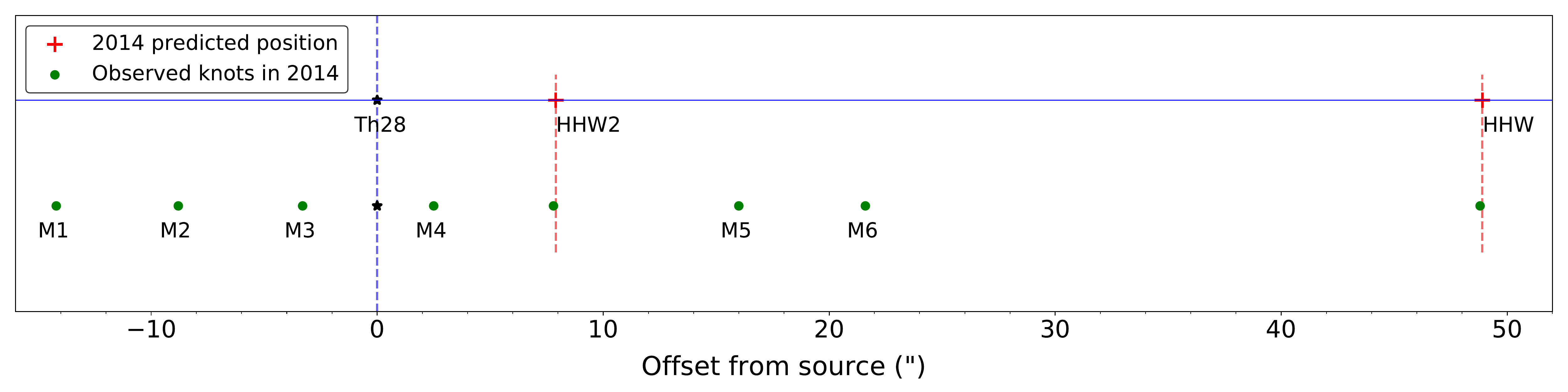}
		\caption{Knot positions along the jet axis within the FOV of our MUSE data (green dots). Also plotted are the predicted positions of HHW and HHW$_{2}$ (red crosses) based on previous velocity and position measurements; these correspond almost perfectly with their positions in our observations.}
		\label{fig:innerknots}     
	\end{figure*}

	Beginning with the blue-shifted jet, the two knots M1 and M2, at projected distances of -14\farcs2 and -8\farcs8 respectively, are well detected in both the deconvolved and unconvolved channel maps and in the flux profiles. M1 is only visible in H$\alpha$ emission and is outside the region of deconvolution. Their bow-shaped morphology is dissimilar from the shape of the other knots. Further in, the knot M3 at -3\farcs3 appears faintly visible as a ‘shoulder’ in the flux before deconvolution, but is a distinct peak in the deconvolved images and flux profiles. To further investigate the veracity of this knot, we examined PV maps from X-Shooter observations of the jet taken in 2015, shown in Figure \ref{fig:xshooter_knots}. These show a knot visible in the [O II] emission of the blue-shifted jet at -3\farcs77.
	
    In the flux profile of the red-shifted jet before deconvolution, a distinct shoulder can be seen at about +2\farcs5 to 3\arcsec\ which becomes a distinct knot (M4) in the deconvolved flux profiles and channel maps. This knot is also seen in channel maps and PV diagrams of the [Fe II]$\lambda$7453 line without deconvolution. The peak is fitted to +2\farcs5 from the source and can also be seen in the X-Shooter data at + 2\farcs85 from the source. The prominent knot at +7\farcs8 is clearly detected in all cases. Additionally, the two weaker knots M5 and M6, at +16\arcsec\ and +21\farcs6, are visible in the flux profiles of H$\alpha$.
	
   The possible knots detected at -6\farcs6 and +5\arcsec\ to +6\arcsec\ in the flux profiles (marked as PK in Figure \ref{fig:axis_fluxprofile}) and in the deconvolved channel maps, may be discussed together since they have several features in common: they appear in similar positions either side of the source, they are both visible as weak bumps in the flux profiles, and while detected in multiple lines and velocity bins, the number of bumps in each region varies, and the spectro-images do not resolve knots in these regions. Perhaps notably, both of these bumpy jet regions lie just behind the large knots M2 and HHW$_{2}$. Given this inconsistency, and that neither the PV diagrams nor the spectro-images before deconvolution show knots at these positions, we consider these to be artifacts of the deconvolution rather than true knot structures.

	\begin{figure*}
		\centering
		\includegraphics[width=18cm, trim= 2.5cm 1.5cm 0cm 0cm, clip=true]{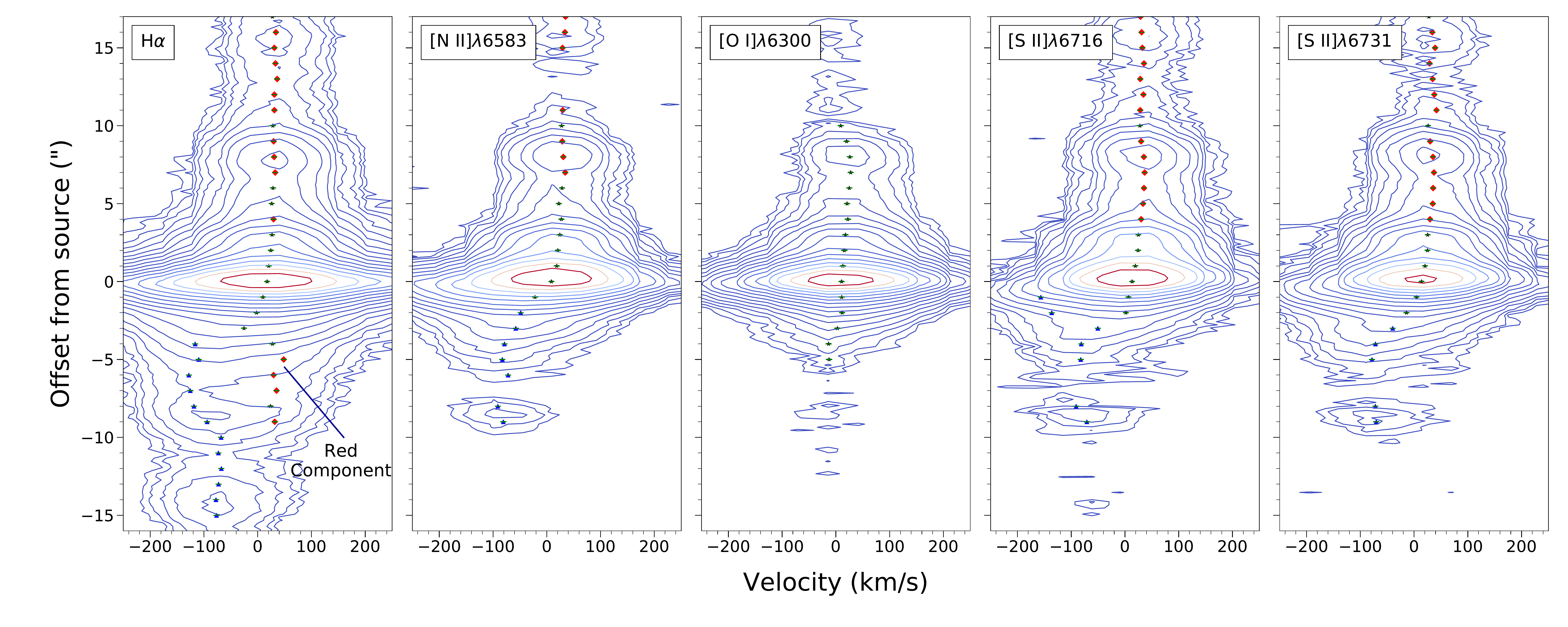}
		\caption{Position-velocity maps of selected jet emission lines with overplotted velocity centroids. Contours begin at the 4-$\sigma$ level, with adjacent contours increasing as a factor of $\sqrt{3}$. Blue and red points indicate blue and red-shifted velocities respectively, while black points indicate centroids within the central channel (approximately +/- 90 km s$^{-1}$). A red-shifted component in the approaching H$\alpha$ jet is annotated.}
		\label{fig:rot_pvmaps}     
	\end{figure*}

	As mentioned, these knots are not reported in earlier studies, with one exception. \cite{Comeron2010} previously identified another knot within this region of the red-shifted jet, designated HHW$_{2}$ and estimate its proper motion at 0\farcs34 $\pm$ 0\farcs05 yr$^{-1}$. Based on the position of this knot at 4\farcs5 in 2004, its position in 2014 would then correspond almost exactly to the bright red-shifted knot centred at 7\farcs8. We therefore identify this as HHW$_{2}$. Similarly, taking the measured position of HHW at 38\arcsec\ in 1985 \citep{Krautter1986} with our measurement of its position at 49\farcs6 in 2014, we also find its proper motion to be 0\farcs40 yr$^{-1}$. Figure \ref{fig:innerknots} shows a comparison plot of all knot positions within the MUSE FOV during the 2014 observations, including the new knots we have identified. We mark both the actual and predicted positions (red crosses) of HHW and HHW$_{2}$. The predicted positions are almost perfectly aligned with their observed positions in 2014, supporting the proper motion estimates previously made.
	
	We calculate the proper motions of the new knots M3 and M4 by comparing their knot positions in the MUSE and X-Shooter observations. (M1, M2, M5 and M6 were not within the spatial range of the X-Shooter data.) This allows us to constrain the proper motions close to the driving source. In the red-shifted micro-jet, we detect M4 at +2\farcs5 in 2014, and at +2\farcs85 in 2015. This gives us an estimated proper motion of 0\farcs35 yr$^{-1}$, consistent with the estimate of \cite{Comeron2010} for this jet. For the blue-shifted lobe, the knot M3 found at $\sim$-3\farcs3 in 2014 is found at -3\farcs77 in 2015, giving an estimated proper motion of 0\farcs47 yr$^{-1}$.  This is the same as the proper motion of the closest blue-shifted HH object HHE$_{1}$, estimated by \cite{Wang09}, and a little higher than the 0\farcs43 yr$^{-1}$~ found by \cite{Comeron2011}.
	
	The improved spatial resolution of the deconvolved jet images make it possible to identify and measure the peak positions of knots M3 and M4 within $\sim$3\arcsec\ of the source and hence to estimate their proper motions. As noted in Section \ref{section:deconvolution}, the sampled PSF used in the deconvolution is slightly elongated along the axis of the jet. However, both knots are verified from the X-Shooter observations, and the measurements based on the deconvolved MUSE data yields proper motion estimates consistent with those found for the other knots in each jet lobe. We therefore consider it unlikely that the elongation of the PSF has a substantial impact on our findings.
	
	Table \ref{table:MUSE_knots} shows a summary of the identified knots, including radial velocities; these are estimated from the data pre-deconvolution, taking a Gaussian fit to the line spectrum summed over a 1\arcsec\ $\times$ 1\arcsec\ spatial region centred on the peak of the knot. Those in the blue-shifted jet show radial velocities of about -60-80~\kms, a factor of 2-3 larger than the +20-35~km s$^{-1}$ radial velocities found in the red-shifted knots; this is consistent with previous estimates for the velocities of HH objects associated with either jet. Table \ref{table:MUSE_knots} also gives the FWHM, proper motions, tangential velocities and inclination angles of the knots. The FWHM values have a typical uncertainty of 3$\%$. For the knots where we could not measure the proper motion the values are assumed to be the same as M3 and M4, for the blue and red lobes respectively. 
	 
    Finally, the estimated ages of the knots are also given in Table \ref{table:MUSE_knots}. We find that several knots appear to be paired with counterparts in the opposite jet. The innermost pair (M3 and M4) at +2\arcsec-3\arcsec\ were both ejected in 2007 ($\pm$ 2 yrs), while HHW$_{2}$ appears to have been ejected in 1992 $\pm$ 7 yrs, close to the first blue-shifted bowshock at -8\farcs8 in 1995 $\pm$ 8 yrs. The blue-shifted bowshock M1, at -14\farcs2, may be the counterpart of the red-shifted knot M5 at +16\arcsec, however their estimated ages and corresponding ejection times (1984 and 1968) are significantly different, even given the large uncertainty of 13 years associated with each. This may be due to the blue-shifted knot being slowed by interaction with the surrounding medium, although we do not have individual proper motion estimates of these knots to compare.


	\begin{table*}
		\centering
		\begin{threeparttable}

		\caption[knot_table]{\label{table:MUSE_knots} Table of knots detected in the MUSE observations. }

		\renewcommand{\arraystretch}{1.3} 

	\begin{tabular}{{p{0.05\textwidth}<{\raggedright} 
			p{0.06\textwidth}<{\raggedright} 
			p{0.08\textwidth}<{\raggedright}
			p{0.1\textwidth}<{\raggedright}
			p{0.07\textwidth}<{\raggedright}
			p{0.07\textwidth}<{\raggedright}
			p{0.07\textwidth}<{\raggedright}
			p{0.09\textwidth}<{\raggedright}
			p{0.11\textwidth}<{\raggedright\arraybackslash}}}
	
	    \hline \hline
	    Jet & Knot &  \makecell[l]{Offset \\ (\arcsec)} &  \makecell[l]{Proper Motion \\ (\arcsec/yr$^{-1}$)} & \makecell[l]{$v_{tan}$ \\ (km~s$^{-1}$)} & \makecell[l]{$v_{rad}$ \\ (km~s$^{-1}$)} & \makecell[l]{FWHM \\     (km~s$^{-1}$)} & \makecell[l]{$i_{knot}$ \\ ($^{\circ}$)} & \makecell[l]{Dynamical Age \\ (yr)} \\
	    \hline
    	Blue & M1 & -14.2 $\pm$ 0.1 & &-366 & -85 & 160 $\pm$ 5 & 13.4 $\pm$ 3.5 & 30.2 $\pm$ 13 
    	\\
    	& M2 & -8.8 $\pm$ 0.1 & &-366 & -80 & 175 $\pm$ 5 & 12.6 $\pm$ 3.5 & 18.7 $\pm$ 8
    	\\
    	 & M3 & -3.3 $\pm$ 0.1 & 0.47 $\pm$ 0.1 &-366 & -60 & 210 $\pm$ 6 & 9.5 $\pm$ 3.0 & 7.0 $\pm$ 2 
    	\\
    	&  &  &  & &
    	\\  
    	Red & M4 & 2.5 $\pm$ 0.1 &0.35 $\pm$ 0.05 & 266 & +20 & 140 $\pm$ 4 & 4.3 $\pm$ 2.5 & 7.1 $\pm$ 2
    	\\ 
    	 & HHW$_{2}$ & 7.8 $\pm$ 0.1 & 0.34 $\pm$ 0.05 & 258  & +30 & 120 $\pm$ 4 & 6.6 $\pm$ 2.5 & 22.3 $\pm$ 7  
    	\\ 
    	 & M5 & 16 $\pm$ 0.1 & &266  & +30 & 120 $\pm$ 4 & 6.4 $\pm$ 2.5 & 45.7 $\pm$ 13 
    	\\
    	 & M6 & 21.6 $\pm$ 0.1 & & 266 & +25 & 120 $\pm$ 4 & 6.4 $\pm$ 2.5 & 61.7 $\pm$ 18 
    	\\ 
    	 & HHW & 49.6 $\pm$ 0.1 & 0.4 $\pm$ 0.05 & 304 & +35 & 120 $\pm$ 4 & 6.6 $\pm$ 2.5 & 124 $\pm$ 15 \\ 
    	\hline  
		
	\end{tabular}

	\begin{tablenotes}
		\small
		\item\textit{Note:} The first column specifies in which jet (blue- or red-shifted) each knot is observed. Distances are projected; radial velocities and FWHMs are measured from the [S II] lines except for M1 which is measured only in H$\alpha$.
	\end{tablenotes} 
	\end{threeparttable}
		
	\end{table*}

    For HHW, with a proper motion of 0\farcs4 $\pm$ 0\farcs05~yr$^{-1}$ \citep{Comeron2010}, we infer it to have been ejected around 1890 $\pm$ 15 years. We compare this with HHE$_{1}$, which was observed at -39\arcsec~ from the source by \cite{Wang09} in their 2004 observations and for which \cite{Comeron2010} gives the ejection year to be 1915 ($\pm$ 7 years, taking the estimated uncertainty of 0\farcs03~yr$^{-1}$ on the proper motion). If this knot represents a counterpart to HHW, it then seems likely that the proper motions of one or both knots have changed significantly over time. 
    
    Previous work estimating the ages of the knots in the extended blue-shifted jet indicated intervals as short as 50 years between ejections, and noted that the decreasing intervals between knots closer to the source might suggest an increase in the frequency of knot ejection. However, as this work shows, Th~28 appears to be forming knots on much shorter timescales of $\sim$10 to 15 years, making it more probable that intervening knots have simply become too faint to be detected; indeed, this can be clearly seen with the red-shifted knots from HHW$_{2}$ to M6, which are progressively fainter between +8\arcsec\ and +22\arcsec\ from the source, with the M6 knot falling below 3-$\sigma$ of the background noise.
	
	\subsubsection{Jet inclination and velocity}
	\label{section:jetvelocity}
	Based on the proper motions and radial velocities of the knots detailed in Table \ref{table:MUSE_knots}, the velocities of the red and blue-shifted jets are estimated at 270 $\pm$ 30 \kms\ and 365 $\pm$ 60 \kms\ respectively, with corresponding inclination angles of +6.5$^{\circ}$ and -13.0$^{\circ}$ to the plane of the sky. We exclude the innermost knots M3 and M4 from this estimate since their radial velocities are likely to be underestimated due to the contributions from scattered light close to the source. Uncertainties on the radial velocities are taken to be 10$\%$ of the velocity resolution, approximately 6 \kms\ (discussed further in Section \ref{subsubsection:kinematics}). 
	
	Our findings are in line with previous estimates based on the proper motions and radial velocities of the outer HH objects (see Section \ref{section:th28_background}). We note that this implies a difference of 5-6$^{\circ}$ in inclination angle between the two jets. Some stellar jets are known to possess lobes with deflected or misaligned axes, and while we detect no difference in the PA of the Th28 micro-jets (discussed in Section \ref{section:precession} below), a small deflection in PA $\leq$ 1$^{\circ}$ cannot be ruled out.

	\subsubsection{Kinematics}
	\label{subsubsection:kinematics}
	
    Figure \ref{fig:rot_pvmaps} shows position-velocity maps of the rotated micro-jets in the main emission lines of interest. The velocity centroids over-plotted are Gaussian-fitted from the spectra binned at 1\arcsec\ intervals along the jet axis, with the uncertainty in velocity taken as 10$\%$ of the velocity resolution ($\approx$ 6-7~km s$^{-1}$). (Note that the standard estimates (Section \ref{section:reduction}) give uncertainties $\leq$ 2~km s$^{-1}$, and we therefore adopt the more conservative measure.) The low peak radial velocities due to the micro-jets' orientation are clearly illustrated here, with emission from the red-shifted jet extending into low-velocity blue-shifted channels. The asymmetry in radial velocities between the jets is also pronounced, with the blue-shifted jet centred at a roughly constant velocity of -80~km s$^{-1}$, greater by a factor of 2-3 than the red-shifted jet at +30~km s$^{-1}$. The bowshocks M1 and M2 are clearly seen in the blue-shifted jet at -8\farcs8 and -14\farcs2, as well as HHW$_{2}$ and M5 at +8\arcsec\ and +16\arcsec\ in the red-shifted jet.
	
	The fitted velocity centroids reveal multiple components in the blue-shifted jet, including red-shifted emission which is the most interesting feature and which we now discuss. Previously \citet{Coffey2007} obtained transverse PV maps of the blue-shifted jet at 0\farcs5 from the source using HST-STIS, and found a double-peaked profile in H$\alpha$ with a low-velocity red-shifted component centred at +30 km s$^{-1}$. The asymmetric profile of this component closely matched that of the red-shifted micro-jet, and it was therefore attributed to contamination of the slit with reflected light from the red lobe. Similarly, our position-velocity maps of the blue-shifted jet before deconvolution also show a red-shifted component (+30-50~km s$^{-1}$); this emission is most prominent in the H$\alpha$ and [S II]$\lambda$6731 lines, and shows a constant velocity along the jet axis. We find that this scattered-light component can be detected to a distance of 9\arcsec, the position of the bowshock M2.

	\begin{figure*}
		\centering
		\includegraphics[width=9cm, trim= 0.45cm 0.5cm 0.6cm 0.8cm, clip=true]{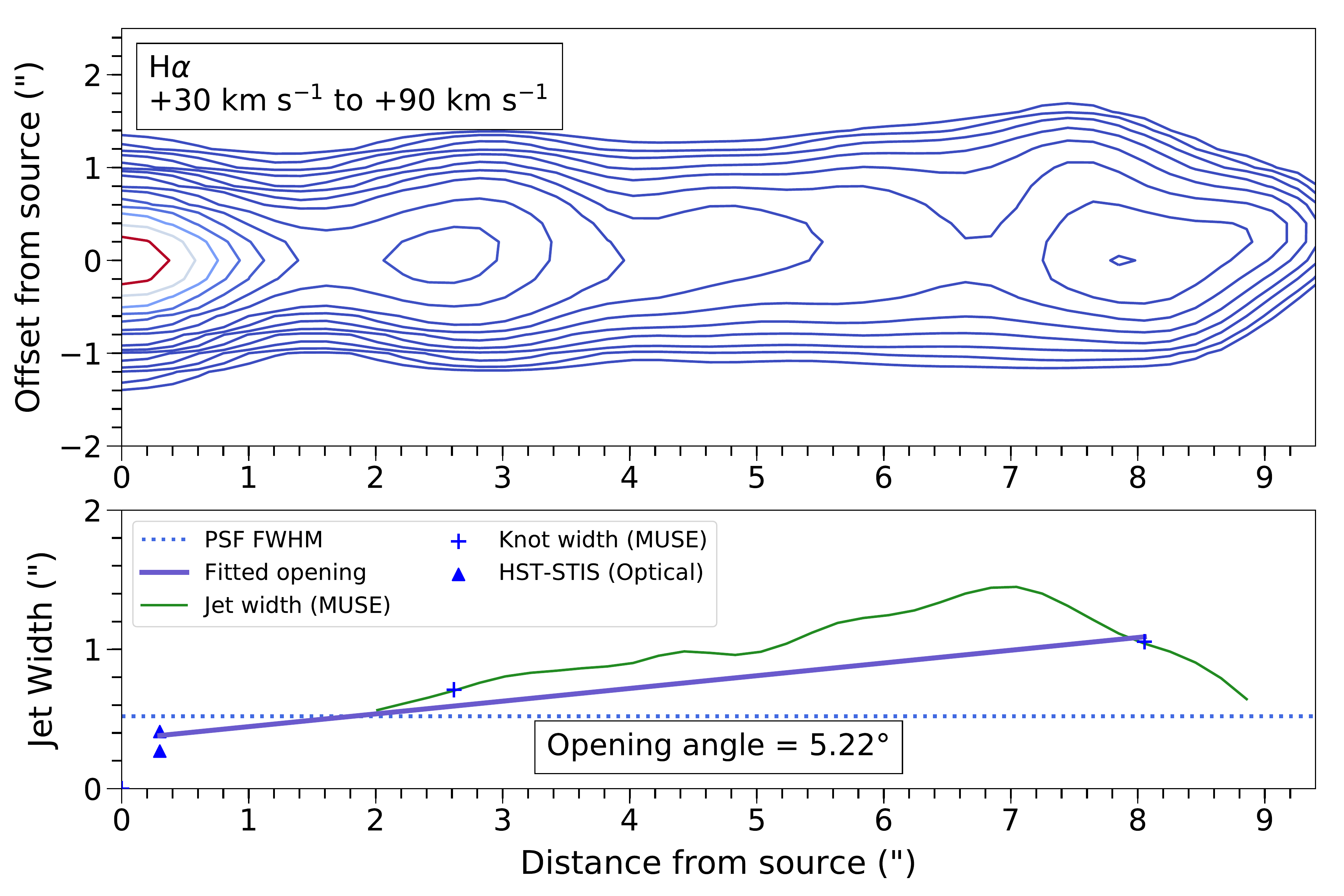}
		\includegraphics[width=9cm, trim= 0.45cm 0.5cm 0.6cm 0.2cm, clip=true]{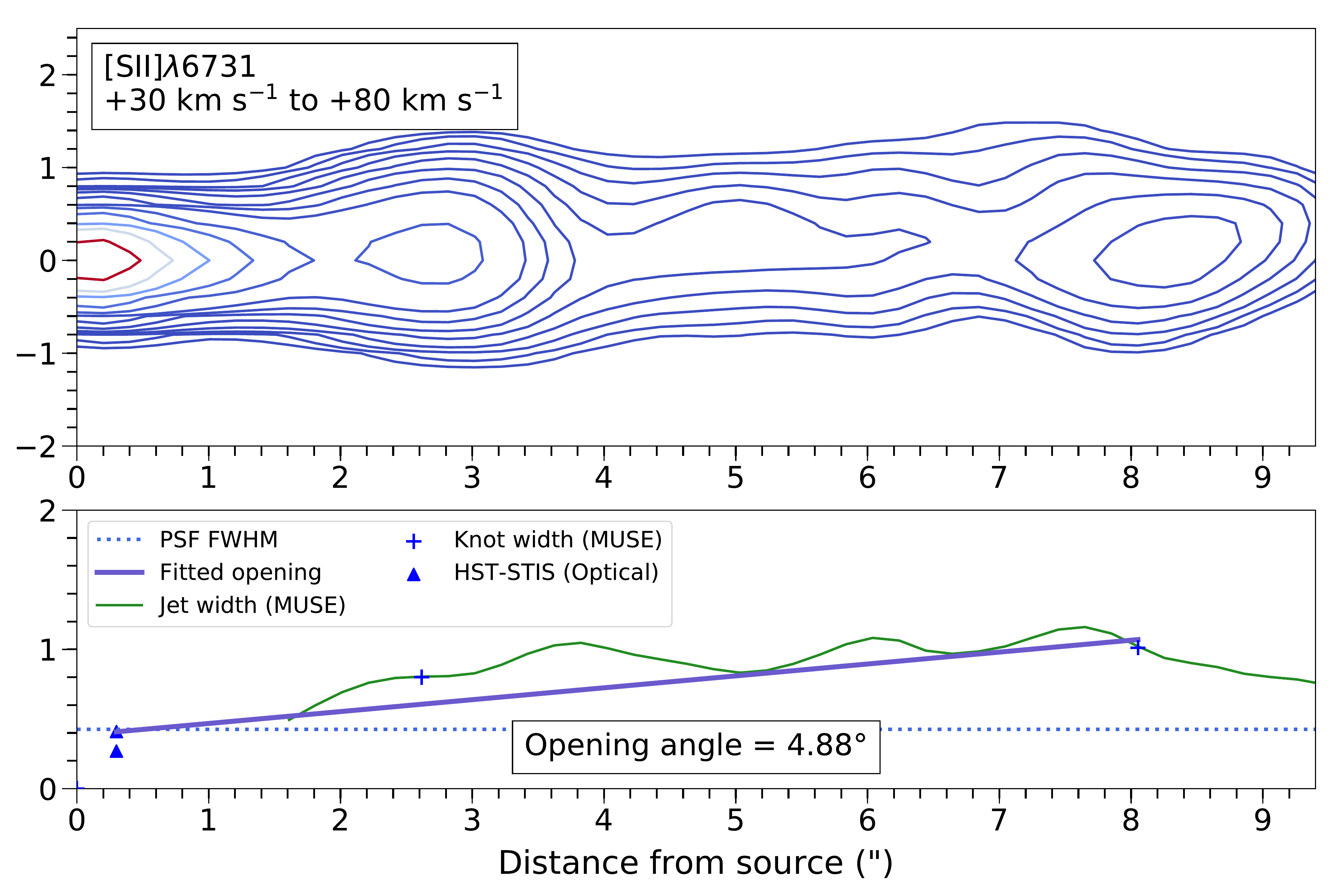}
		\caption{Lower panels: Intrinsic jet FWHM (plotted in green) measured along the red-shifted micro-jet, with the source position at the origin. Deprojected distance from the source is given on the x-axis. The opening angle in each case is estimated by a linear fit to the lower bounds of the FWHM measurements, corresponding approximately to observed knot positions (marked with blue points). Upper panels show corresponding channel maps of the deconvolved channel maps; contours begin at 3$\sigma$ of the background noise and increase as a factor of $\sqrt{3}$. }
		\label{fig:jetwidth_red}     
	\end{figure*}
	
	\begin{figure*}
		\centering
		\includegraphics[width=9cm, trim= 0.45cm 0.5cm 0.6cm 0cm, clip=true]{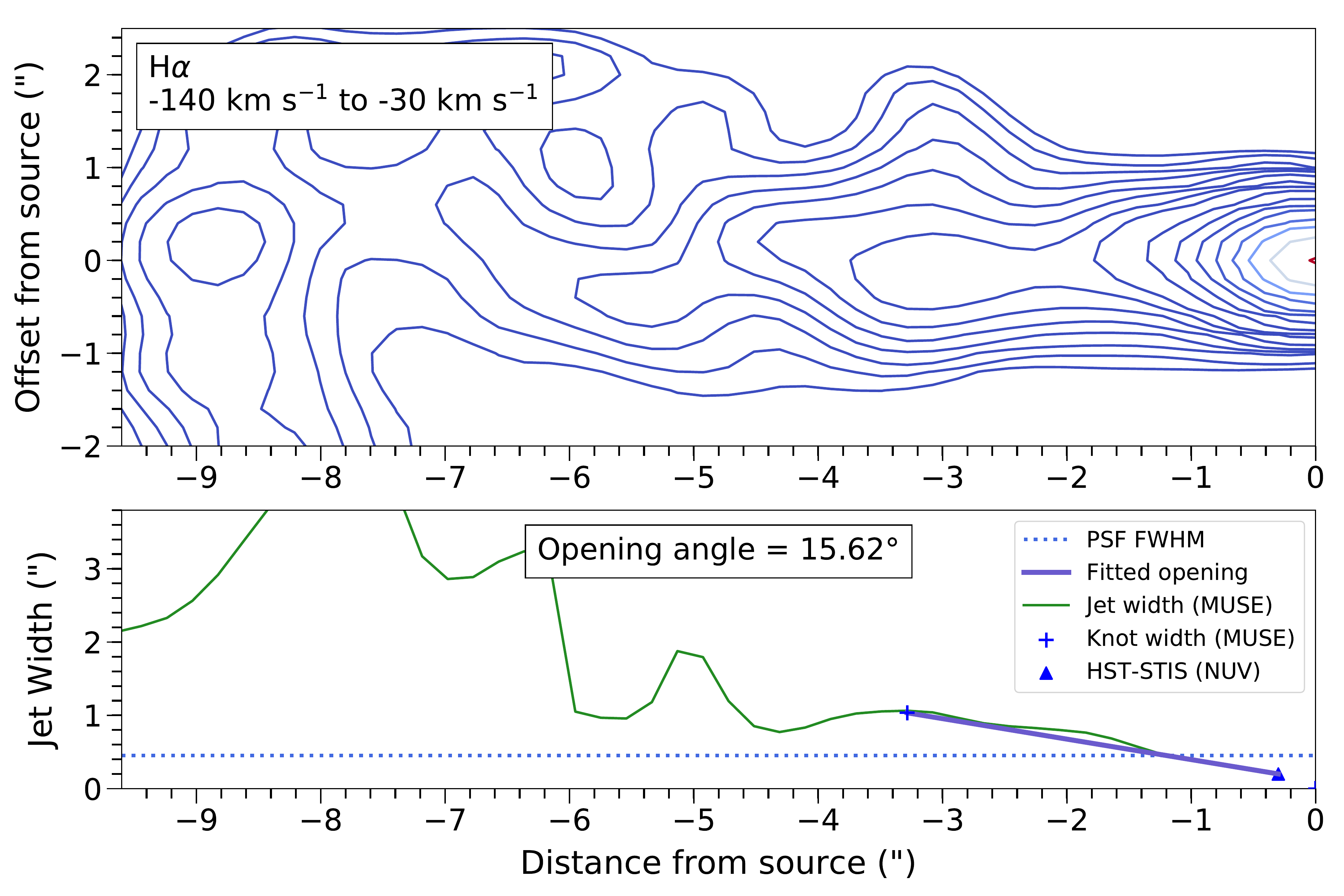}
		\includegraphics[width=9cm, trim= 0.35cm 0.5cm 0.6cm 0.2cm, clip=true]{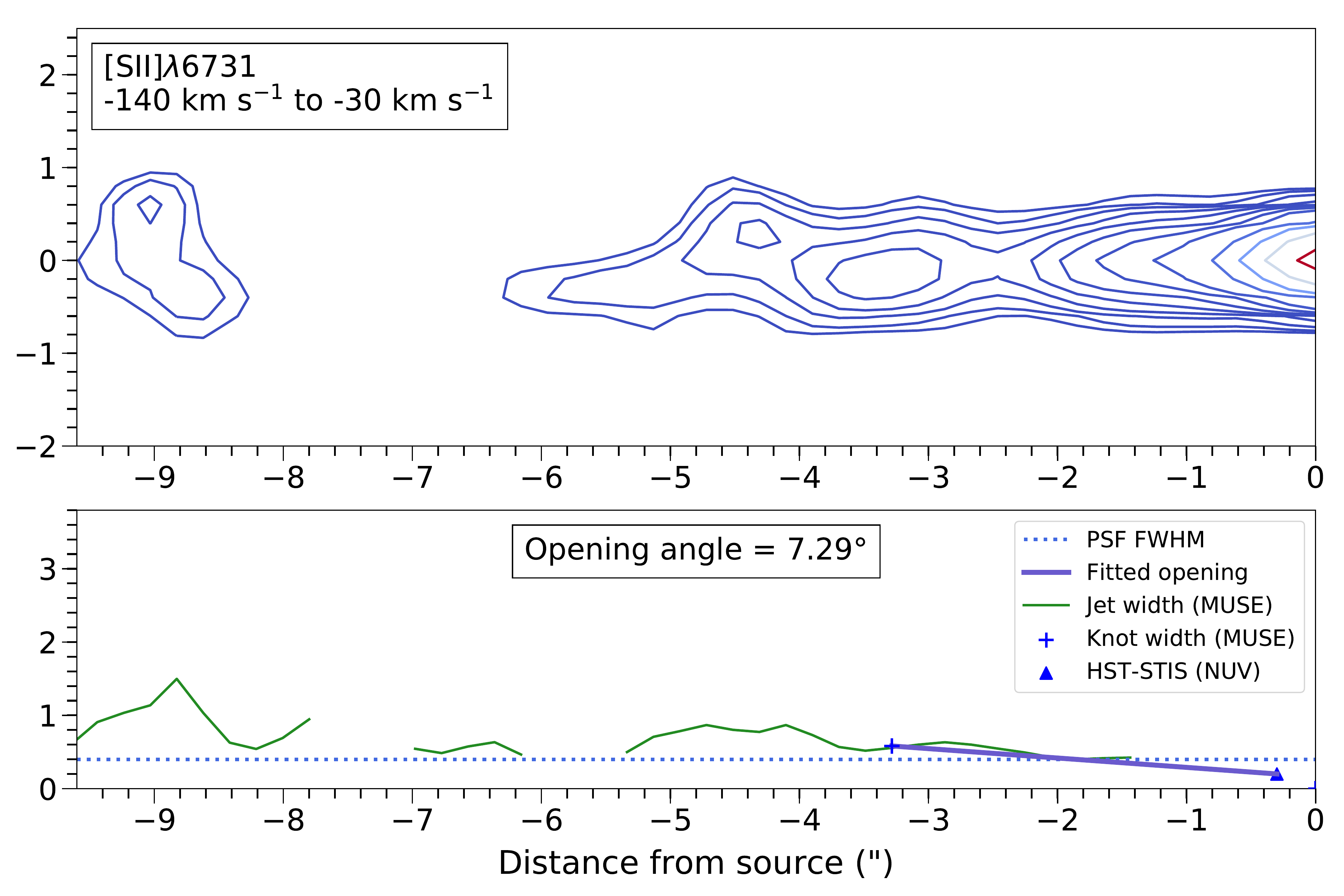}
		\caption{Lower panels: Intrinsic jet FWHM measured along the blue-shifted micro-jet, with the source position at the origin. Distance from the source is deprojected. The opening angle in each case is estimated by a fit to the lower bounds of the FWHM measured for the main peak at the observed knot positions. Note that the plot in green corresponds to the overall width measured by fitting a single Gaussian across the jet and is therefore substantially wider where a bow shock is present in H$\alpha$. Upper panels show corresponding channel maps of the deconvolved channel maps; contours begin at 3$\sigma$ of the background noise and increase as a factor of $\sqrt{3}$.}
		\label{fig:jetwidth_blue}     
	\end{figure*}
	
    In addition to the position-velocity maps taken parallel to the jet axis, we examined maps of transverse cuts across the jet, taken at knot positions across both micro-jets. A velocity shift of 10~km s$^{-1}$ - 20~km s$^{-1}$ across the red-shifted jet (in the southwest-northeast direction), suggesting rotation around the axis, was previously observed by \cite{Coffey2004} at $<$ 0\farcs3 ($\sim$48~au) from the source. In fitting the velocity centroids of our transverse maps we observe no such trend in either jet. This is unsurprising, both due to the significantly lower velocity resolution of MUSE and the difference in spatial resolution. Detecting such rotation requires resolving the jet width close to the base, before this signature is lost due to other kinematic features. In these seeing-limited observations, the jet is not resolved until approximately 4\arcsec\ from the source, at which distance we clearly detect signatures of kinematic features such as shocks and wiggling, which have a high likelihood of contaminating any rotation signatures. However, future MUSE observations with adaptive optics corrections would have the potential to reveal jet rotation signatures in optical lines.

	\subsection{Jet width}
	
    The jet width is measured from the deconvolved data cubes by fitting a Gaussian function to the jet cross-section, centred at at the peaks of the detected knots (avoiding the wings of the spatial PSF in the x-direction) to obtain accurate estimates of the jet width \citep{Raga1991, Mundt1991}. Each cross-section has spatial binning of 1\arcsec\ along the jet axis.  The jet width at the source position is subtracted in quadrature from the fitted width along the jet axis to recover the intrinsic width. Examples of the resulting fits for both micro-jets are shown in Figures \ref{fig:jetwidth_red} and \ref{fig:jetwidth_blue} (lower panels). Measurements below the FWHM of the source PSF were excluded. To estimate the full jet opening angle $\alpha$ we combine these with the HST-STIS measurements from \citet{Coffey2007}. In the red-shifted jet this adds optical-line jet width estimates of 0\farcs27 and 0\farcs41 (at +0\farcs3 from the source), with NUV measurements for the blue-shifted jet giving an upper estimate of 0\farcs2 (-0\farcs3 from the source). Note that for these observations no contribution was found from the stellar PSF and so we do not subtract this. For each emission line and velocity bin, the opening angle is calculated using the deprojected knot positions (with \textit{$i_{red}$} = 6.5$^{\circ}$, \textit{$i_{blue}$} = -13.0$^{\circ}$) and we quote the positions below accordingly.
    
    Table \ref{table:open_angles2} shows the estimated full opening angles for both jets based on several velocity bins of the H$\alpha$ and [S II]$\lambda\lambda$6716, 6731 emission. In the red-shifted jet we measure an average width of 130 au at knot M4 (+2.6\arcsec), with a slow increase to 170 au at the position of HHW$_{2}$ (+8\arcsec). For the red-shifted jet we find opening angles of 4.5-6.7$^{\circ}$, with no clear trend between velocity channels. These results are in line with opening angles of 2$^{\circ}$ - 4$^{\circ}$ measured in a number of other jets \citep{Burrows96, Dougados2000}. 
    
    In the case of the blue-shifted jet, we measure the width only at the knot M3 (-3\farcs5). As described above, the more distant knot M2 has a bow-shock morphology which is not representative of the jet width. As this jet is relatively faint, larger velocity bins were required to obtain a good fit. We see a significant difference in the results between emission lines. In H$\alpha$, M3 has a width of 165 au ($\alpha$ = 15.6$^{\circ}$) whereas in both [S II] lines it is only 90-115 au ($\alpha$ = 7-10$^{\circ}$). This suggests that the H$\alpha$ emission traces a wider component of the flow; such a large discrepancy may further indicate that M3 is an unresolved bow shock, with H$\alpha$ emission tracing the shock wings and [S II] emission limited to the central peak.

	\begin{table}
		\centering
		
		\caption[openangles]{Table of fitted opening angles for both micro-jets based on deconvolved H$\alpha$ and [S II] emission. 	}
		
		\begin{tabular}{{p{0.1\textwidth}<{\raggedright} p{0.1\textwidth}<{\raggedright} p{0.05\textwidth}<{\raggedright} p{0.06\textwidth}<{\raggedright}  p{0.03\textwidth}<{\raggedright\arraybackslash}}}
			\hline \hline
			Jet & Line & \makecell[l]{$>$ $v_{rad}$ \\(\kms)} & \makecell[l]{$<$ $v_{rad}$ \\(\kms)} & \makecell[l]{$\alpha$ \\ ($^{\circ}$)} \\ 
			\hline 
			&  & &  &  \\ 
			Red-shifted & H$\alpha$ & -30 & +30 & 4.8 \\ 
			
			& & +30 & +90 & 5.2 \\ 
			
			& & +90 & +150 & 4.4 \\ 
			& & -30 & +150 & 4.9 \\ 
			&  & &  &  \\ 
			
			& [S II]$\lambda$6716 & -30 & +30 & 6.7 \\ 
			
			& & +30 & +80 & 6.2 \\ 
			
			& & +80 & +140 & 5.5 \\ 
			& & -30 & +140 & 6.2 \\ 
			& &  &  &  \\ 
			& \rule[-1ex]{0pt}{2.5ex}[S II]$\lambda$6731 & -30 & +30 & 4.5 \\ 
			
			& & +30 & +80 & 4.9 \\ 
			
			& & +80 & +140 & 4.8 \\ 
			& & -30 & +140 & 4.8 \\ 
			
			& &  &  &  \\
			
			Blue-shifted & H$\alpha$ & - 30 & - 140 & 15.6 \\ 
			& &  &  &  \\ 				
			& [S II]$\lambda$6716 & - 30 & - 140 & 9.9 \\
			& &  &  &  \\ 				
			& [S II]$\lambda$6731 & - 30 & - 140 & 7.3 \\
			
			\hline 
		\end{tabular} 

		\label{table:open_angles2}
	\end{table}
	
	\subsection{Jet centroid}
	\label{subsection:centroids}
	
\begin{figure}
		\centering
		\includegraphics[width=8.6cm, trim= 1cm 0.2cm 0.3cm 0.1cm]{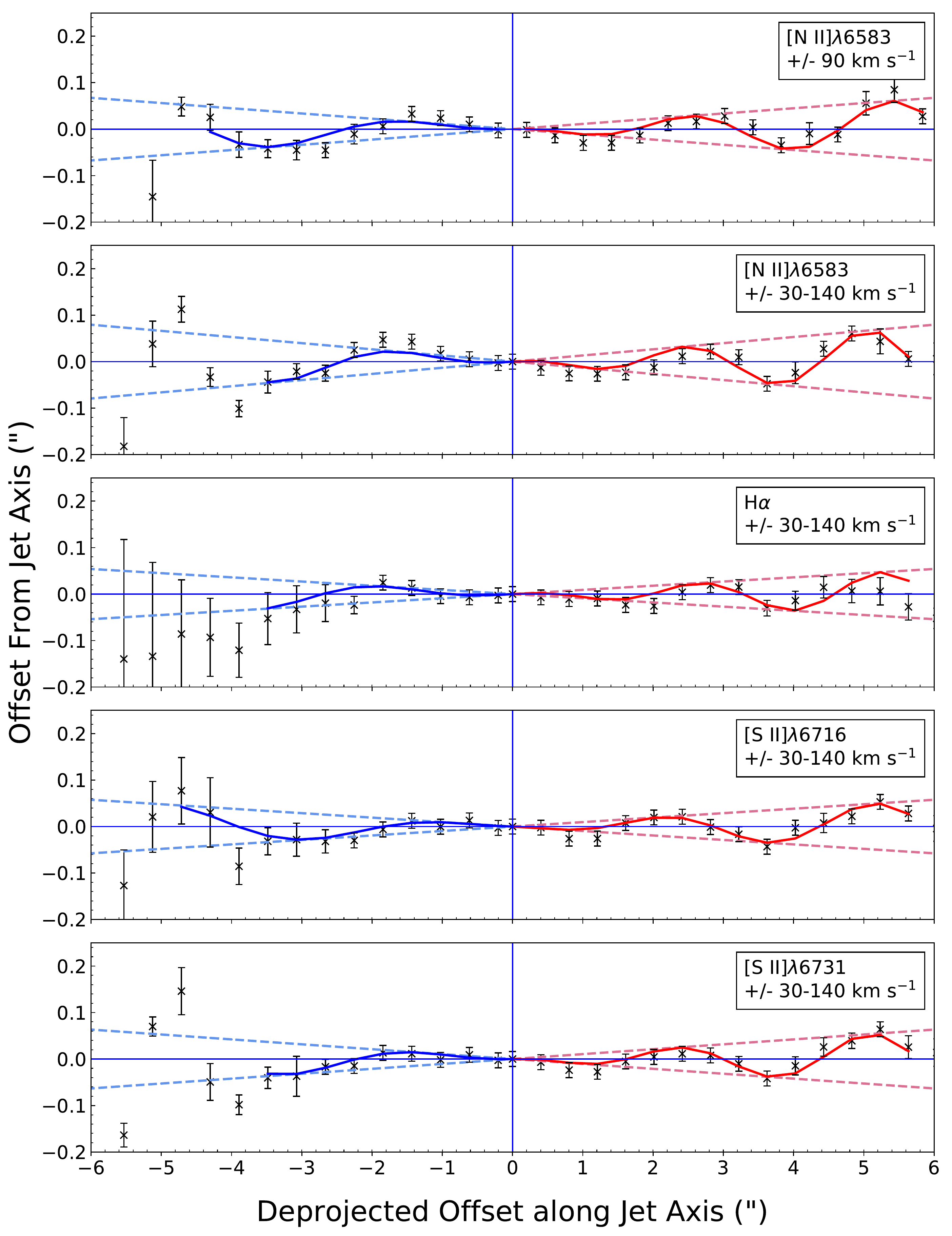}
		\caption{Precession models fitted to the centroids of the blue- (left) and red- (right) shifted lobes of the micro-jets on either side of the jet source at 0'', shown for several emission lines. The [N II]$\lambda$ centroids in the top panel are sampled from +/- 90 ~\kms\ in both lobes; for all other lines the blue-shifted (-30 to -140 ~\kms) and red-shifted (+30 to +140 ~\kms) lobes are sampled separately. The dashed lines mark the bounds of the successive maxima and minima defined by the half-opening angle $\beta$ in each case.}
		\label{fig:wigglefit}     
	\end{figure}

	As well as using Gaussian fitting to study the jet width, the position of the jet centroid was investigated as a function of distance from the driving source. The same fitting procedure as described above was adopted. Both jets show indications of displacement around the jet axis. In particular, we find a wiggling pattern with a clear point symmetry within approximately 6\arcsec\ of the source. Figure \ref{fig:wigglefit} presents the estimated jet centroid positions (relative to the source x and y position) for selected emission lines before deconvolution. 
	
	The uncertainty on the centroids was calculated using the standard error (see Section \ref{section:reduction}); however, for centroids at positions close to the source the strong line emission around the star gives extremely high SNR ratios ($\geq$ 1000) and correspondingly small error bars. This is unlikely to represent the true accuracy of the estimated jet position in this region due to the unknown contributions from, e.g., scattered light and unresolved shocks. We therefore estimate a minimum uncertainty by looking at the scatter between the centroid positions measured in different emission lines. We find that for centroids measured at the same spatial position within $\pm$ 1\arcsec\ of the source, the average standard deviation is 0\farcs012. We therefore adopt this as the minimum uncertainty of the centroids.
	
\begin{figure*}
		\centering
		\includegraphics[width=18 cm, trim= 1.5cm 1cm 0.5cm 0cm, clip=true]{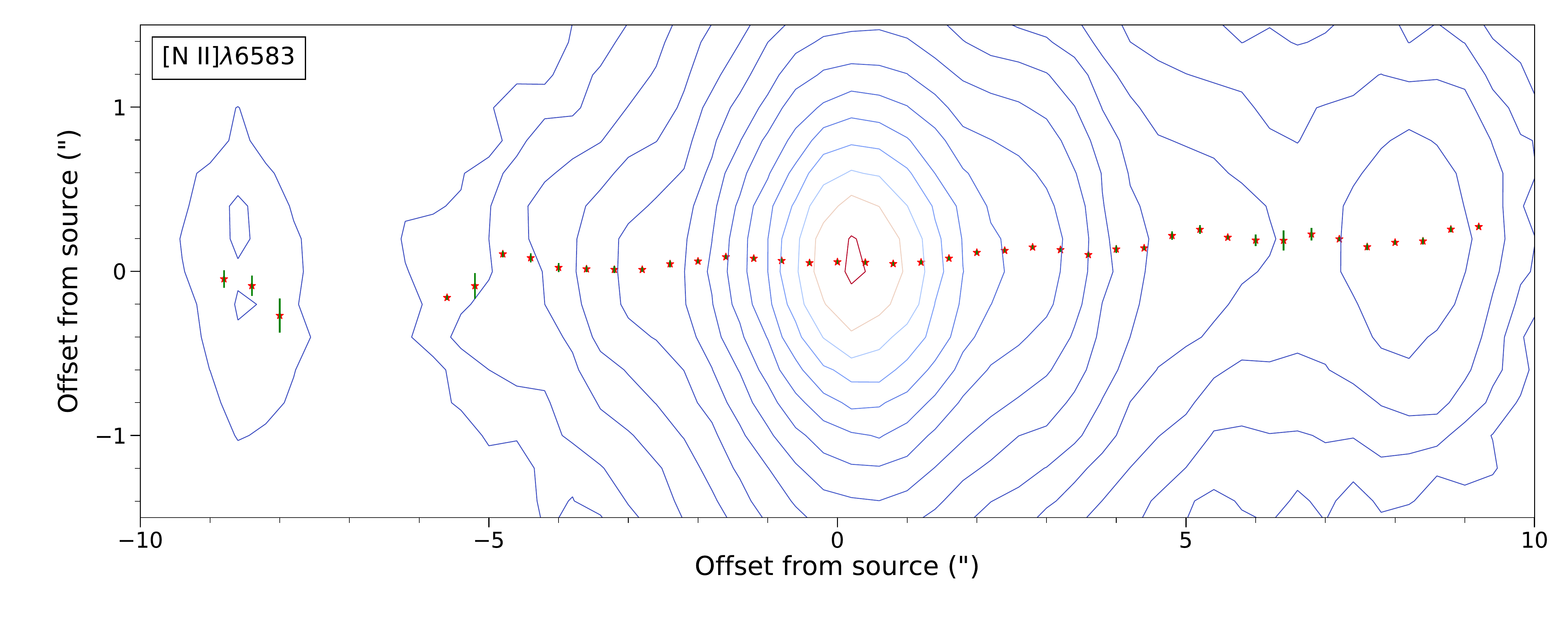}
		\caption{Centroid positions fitted to the inner jet in the brightest velocity channel of [N II] (+/- 90 \kms). Contours start at 12-$\sigma$ of the background noise (60 $\times$ 10$^{-20}$ erg~s$^{-1}$~cm$^{-2}$ $\AA^{-1}$) and increase as factors of $\sqrt{3}$.}
		\label{fig:centroid_overlays}     
	\end{figure*}

	We find that the wiggle pattern is most clearly defined in the [N II] centroids; the same pattern can also be seen in the H$\alpha$ and [S II] jets when the lowest velocity channel (+/- 30 \kms) is excluded, as these lines suffer from contamination by low-velocity emission from outer layers of the jet, while the [N II] emission traces the flow closest to the jet axis. This agreement suggests that the wiggle observed is a real feature of the data. The wiggles appear to have a maximum amplitude of 0\farcs1 - 0\farcs2 from the jet axis, corresponding to between half a pixel and one pixel in the field of view. In Figure \ref{fig:centroid_overlays} the centroid positions fitted to the  red-shifted jet in the brightest velocity channel of [N II] are presented overlain on the corresponding channel map, showing how the wandering in the centroid positions follows the contours of the jet
	
	\section{Discussion}

	\subsection{Asymmetries}
	
	Th~28 exhibits a number of marked asymmetries between the jet lobes in terms of morphology and kinematics \citep{Wang09, Comeron2010}. The detection of emission from molecular lines only in the blue-shifted jet also indicates a significant difference in the extinction or excitation between the jet lobes \citep{Coffey2010}. The MUSE observations show in detail the different morphologies of these lobes. The bright red-shifted jet shows an extended well-collimated flow with a series of compact knots; the blue-shifted jet, in contrast, appears continuous only for $\sim$5\arcsec\, and the knots it contains form wide bow shocks with multiple peaks. The blue jet is also much fainter overall, with a substantial degree of diffuse H$\alpha$ emission which may stem from contamination by the extended wings of the bow shocks. Consequently, although the blue-shifted jet appears to be broader and less collimated than the red-shifted jet, the width is difficult to measure, and the appearance of a very uncollimated jet may be in part  due to these unresolved bow shocks within the jet. 
	
	Similar to other studies, we also observe strong disparities in the kinematics of the jets. The red-shifted jet again presents a relatively simple picture, with a single velocity component at $\sim$ +30 \kms\ remaining consistent along the jet axis, while in the blue-shifted jet, we find markedly higher radial velocities (-80 \kms) by a factor of 2-3. Recent high-resolution observations of these micro-jets with HST/STIS show that this asymmetry can be detected within a few tens of au from the star itself \citep{Liu2020}, which may suggest that the asymmetry originates with the launching velocity of the jets. Studies have noted differences between the inclination angles of the two jets estimated from combining the proper motion and radial velocity measurements \citep{Comeron2010, Louvet2016}, and this is also borne out by the results presented in Table 2.
	
	Although we are presenting our diagnostic study in a later paper, we note here that we find only the blue-shifted jet detected in [O III]$\lambda$ 5007 (Figure \ref{fig:main_specims}) whereas He I and emission lines from refractory species ([Fe II], [Ni II], [Ca II]) are found only in the red-shifted jet (Appendix Figures \ref{fig:spec_cat1} to \ref{fig:spec_cat5}). The detection of [O III]$\lambda$ 5007 is of particular significance, as this is an indicator of strong shocks with $v_{shock} >$ 100 \kms: this emission is clearly detected in the peak of the M2 bowshock as well as within the inner 5\arcsec\ of the blue-shifted jet.

	The overall picture of a faster, fainter, more ionised blue-shifted jet contrasting a brighter, slower red-shifted jet is borne out by these findings, including the estimated proper motions of the inner knots M3 and M4. However, observations with higher spatial and spectral resolution would be required to disentangle how much of the morphological and kinematic asymmetry between the two jets is intrinsic at launch, as opposed to the result of the blue-shifted flow interacting with a significantly different environment.

	\subsection{Precession in the inner jet}
	\label{section:precession}
	
    Wiggles similar to that observed in the initial Th~28 jet channel have been observed in other T Tauri micro-jets \citep[e.g.][]{Masciadri2002}. The primary explanation for this is a variation in the direction of jet launching due to either orbital motion of the source due to a companion, or precession of the accretion disk \citep{Terquem1999}. Such precession may also indicate the presence of a companion orbiting on a plane inclined at an angle to the disk, in which case it is expected to coexist with the effect of orbital motion. These two processes can produce similar wiggling morphology, but can be distinguished by the symmetry of the wiggle on either side of the jet source; as discussed in the above paper, orbital motion is expected to produce a jet with a mirror-symmetric wiggle, whereas precession would result in a wiggle with point symmetry across the origin. An S-shaped symmetry has been detected in the knots of several jets at comparatively large distances from the source (200\arcsec\ to 2\arcmin), for example in PV Cep/HH 315 \citep{Gomez1997}, L1157 \citep{Gueth1996,Podio2016}, IRS 54 \citep{Khanzadyan2004, Garcia2013} and IRAS 20126-4104 \citep{Shepherd2000, Caratti2008}. These wiggling jets are typically observed in molecular near-infrared and radio emission and show precession periods on the order of several hundred or thousand years.
    
    However, precession signatures at large distances from the jet are likely to be significantly altered due to the effects of drag, shocks and interactions between knots within the jet. In order to constrain the original wiggle shape (particularly for wiggles with small precession periods and/or opening angles) it is ideal to study the inner jet. In most cases where a wiggle is observed on small scales ($<$ 100\arcsec\ from the source), only one lobe is observed and therefore the wiggle symmetry cannot be detected. We find only two other cases similar to Th~28 where the symmetry of a wiggling jet is identified close to the source. HH 211 \citep{Lee2009,Lee2010}, is a jet oriented close to the plane of the sky with a wiggle extending to approximately 15\arcsec\ in both jet lobes, with a small half-opening angle (0.55$^{\circ}$) and wiggle length scale ~5\farcs5. Second, HH 111 presents an interesting case in which the symmetry of a wiggling jet with a small half-opening angle (0.75$^{\circ}$) can again be traced close to the source. In this case, however, the wiggle extends to a larger length scale $\sim$216\arcsec\ similar to the examples of precessing jets. In both of these examples, however, the jet shows a reflection symmetry around the source indicating the more likely cause is orbital motion, the opposite case to that of Th~28. The clear point symmetry visible in this case leads us to focus on the precession scenario.

	We therefore fitted the centroid positions of the inner jet ($<$ 4-6\arcsec) with the model for a precessing ballistic jet given by \cite{Masciadri2002}. Here the offset of the jet centre from the axis is given by 
	\begin{equation}
	\begin{aligned}
	y = x\tan{\beta} \cos{\left[\dfrac{2\pi }{\tau_{p}}\left(t - \dfrac{x}{v_{j} \cos{\beta}}\right)\right]}
	\end{aligned}
	\end{equation} 
	
	where $\beta$ is the half-opening angle of the precession cone. The jet velocity $v_{j}$ is a key parameter and we therefore attempt to better constrain this for the inner regions of the two micro-jets, by using the deprojected jet velocities estimated for - each side in Section \ref{subsubsection:newknots}. We therefore fix $v_{j, red}$ = 270 \kms\  and $v_{j, blue}$ =  360 \kms\ (Section \ref{section:jetvelocity}) with jet inclination \textit{$i_{red}$} = 6.5$^{\circ}$, \textit{$i_{blue}$} = -13.0$^{\circ}$.
	
	To account for the difference in jet velocity between each jet lobe, we use a combined model comprising two precession components with the same period, precession angle and phase. We then fitted this model to the [N II] centroids along the deprojected jet axes. The fit is weighted using the uncertainty on the centroids described in Section \ref{subsection:centroids}.

	As the rotation did not align the jet axis perfectly at 0$^{\circ}$, we correct for a small residual slope by also fitting a linear component in the model which is then subtracted from the data. The same correction is applied to each lobe, with no significant difference found in the PA of the red and blue-shifted micro-jets. This residual angular offset was found to be $\approx$1.5$^{\circ}$ and combined with the estimated angle in Section \ref{section:reduction} to give a total estimated PA of -83.3$^{\circ}$ and +96.7$^{\circ}$ for the red and blue micro-jets respectively. This is compatible with the PA estimated by \citet{Graham88} as well as the PA of 97.3$^{\circ}$ found for the disk rotation axis by \citet{Louvet2016}.

	Figure \ref{fig:wigglefit} shows the fitted wiggles in several emission lines, with the results given in Table \ref{table:prec_table2b}. $\chi^{2}_{red}$ values for the fits vary betweeen 0.6-1.8.  We find average $\beta$ = 0.6$^{\circ}$, with $\tau_{p}$ of 7.9  years. The time $t$ functions as a phase parameter with an average value of 7.0 years. This parameter varies significantly in the fits, which is likely to reflect in part the added spatial uncertainty along the jet due to the seeing and the subsequent variation in the estimated length scale of the wiggling, $\lambda$. 

	We note also that the blue-shifted lobe is generally less well matched to the fits. This is due to the relative faintness of the blue-shifted jet (reflected in the higher uncertainties of the spatial centroids), combined with the longer length scale of the wiggling in this lobe. In the precessing jet model, it is expected that the length scale $\lambda$ of the jet wiggles will scale with the jet velocity. In the case of Th~28 we then expect this will be larger by a factor of 1.3 in the blue-shifted jet. With $\lambda_{red}$ $\sim$2.8\arcsec\ this then gives $\lambda_{blue}$ $\sim$3\farcs5. As this lobe is only resolved to $\sim$4\arcsec, it is likely that we are less able to accurately fit the precession parameters in this jet.

	Additionally, this model predicts that the radial velocity of a precessing jet should follow a sinusoidal pattern with amplitude $v_{rad, max} = v_{j}\sin{\beta} \cos{\phi}$, with phase such that $ v_{rad}$ is maximum where the spatial offset from the jet axis is 0. Taking the fitted parameters for the red-shifted jet ($\beta$ =0.6$^{\circ}$) this gives an amplitude of $<$ 3 \kms\ using the jet velocity of 260 \kms\ estimated from proper motions ($\phi=6.5$$^{\circ}$). We therefore compared the radial velocity and spatial centroids in [N II], H$\alpha$ and [S II] in the red-shifted jet and attempted to fit a radial velocity wiggle with the form and phase described. Although the radial velocities in the 0''-6'' range of the red-shifted micro-jet show variations on a scale of +/- 1-5 \kms, the model was a poor fit to the data with $\chi^{2} \geq$ 7.0 This is not surprising as such a velocity shift is significantly smaller than the velocity resolution of MUSE, and is therefore unlikely to be detected in these observations.

	\begin{table*}
	\caption[prec_table]{Table of fitted precession parameters for the red and blue-shifted lobes of the Th~28 micro-jets.}
	\label{table:prec_table2b}
	\renewcommand{\arraystretch}{1.5}
	\centering
	\begin{tabular}{p{0.09\textwidth}<{\raggedright} p{0.15\textwidth}<{\raggedright} p{0.03\textwidth}<{\raggedright}	p{0.03\textwidth}<{\raggedright} p{0.03\textwidth}<{\raggedright} p{0.05\textwidth}<{\raggedright} p{0.05\textwidth}<{\raggedright} p{0.05\textwidth}<{\raggedright} p{0.05\textwidth}<{\raggedright} p{0.05\textwidth}<{\raggedright} p{0.04\textwidth}<{\raggedright\arraybackslash}} 
		\hline
		Line     & \makecell[l]{Velocity bin \\ (\kms)} & \makecell[l]{$\chi^{2}_{red}$}  &  \makecell[l]{$\beta$ \\ ($^{\circ}$)} &\makecell[l]{$\tau_{p}$\\ (yr)~} & \makecell[l]{\textit{t} \\ (yr)} & \makecell[l]{{$v_{o,max}$} \\ (\kms) } & \makecell[l]{$\mu$}   & \makecell[l]{ $M_{c}$ \\ ($M_{Jup}$) } &\makecell[l]{$\tau_{o}$\\ (days)~} & \makecell[l]{\textit{a} \\ au }  \\
		\hline
		[N II]$\lambda$6583  & $\pm$ 90 \kms      & 1.4 & 0.6 & 8.2 & 7.1 & 3.0 & 0.045 & 75 & 52 & 0.32 \\
		                     & $\pm$(30-140 \kms) & 1.8 & 0.8 & 7.4 & 6.9 & 3.5 & 0.054 & 90 & 56 & 0.33 \\
		H$\alpha$            & $\pm$(30-140 \kms) & 1.0 & 0.5 & 7.4 & 7.8 & 2.4 & 0.03  & 51 & 31 & 0.23 \\\relax
		[S II]$\lambda$6716  & $\pm$(30-140 \kms) & 0.9 & 0.6 & 8.6 & 6.0 & 2.6 & 0.035 & 58 & 42 & 0.28 \\\relax
		[S II]$\lambda$6731  & $\pm$(30-140 \kms) & 0.6 & 0.6 & 7.8 & 6.6 & 2.8 & 0.039 & 66 & 43 & 0.28 \\
		\hline
	\end{tabular}
\end{table*}

	\subsection{Could Th~28 possess a companion?} 
	
	Could the precession observed be due to a companion orbiting on an inclined plane to the disk? Recent L-band observations of the disk with NAOS-CONICA (NACO) rule out signatures of a stellar-mass or brown dwarf companion at distances greater than 20 au (F. Comer{\'o}n, personal comm., 2020). In this section we therefore examine the possibility of a companion in a very close orbit, resulting in a broken inner disk region precessing around the star \citep{Zhu2019}. Based on the very small angle of precession ( $\leq$ 1$^{\circ}$) this inner disk region would have a correspondingly small inclination to the outer disk.
	
	Although the shape of the jet caused by precession cannot be used to directly derive the properties of such a companion (aside from the inclination angle of the orbital plane), we can use it to constrain the possible mass and orbit of the companion. This is done using the theoretical relationship between the wiggles in the jet shape due to precession, and that caused by the orbital motion of the companion; this relationship is discussed in \cite{Masciadri2002} \citep[see also][]{Anglada2007}.

	Our first constraint is introduced by the fact that the opening angle of the jet due to orbital motion ($\tan{\kappa} = v_{o}/v_{j}$) should be smaller than the opening angle of precession $\beta$, giving us the condition $v_{o}/v_{j} < \tan{\beta}$. Table \ref{table:prec_table2b} thus shows the maximum orbital velocities based on the estimated jet velocity and opening angle found by each fit; these range between 2-7 km s$^{-1}$. Additionally, in most systems the precession period is expected to be at least a factor 10 greater than the orbital period. From \cite{Anglada2007}, the ratio between precession and orbital periods $\tau_{p}$ and $\tau_{o}$ is given by:
	
	\begin{equation}
		\begin{aligned}
			\dfrac{\tau_{o}}{\tau_{p}} = \dfrac{15}{32} \dfrac{\mu}{\sqrt{1 - \mu}} \sigma^{3/2} \cos{\beta} 
		\end{aligned}
	\end{equation}
	
	with $\mu$ the ratio of the companion mass $M_{c}$ to the total mass of the system ($M_{sys}$), and $\sigma = R/a$ the ratio of the circumstellar disk radius \textit{R} to the binary separation \textit{a}. In the case of a stellar mass companion, the circumstellar disk is expected to be truncated by the tidal interaction with the companion such that $\frac{1}{4} \leq \sigma \leq \frac{1}{2}$ \citep{Terquem1999}. For our results, however, the small precession period ($<$ 10 years) implies very small orbital periods ($<$ 1 year) and binary separations $<$ 1 au. Such a close orbit is likely to result in a broken disk, with an inner disk region precessing around the star \citep{Zhu2019}. This disk breaking scenario has been proposed for several systems, including as a source of the precession observed in the RY Tau jet \citep{Garufi2019}. A modified form of the above relation then relates the orbital period to the precession of the inner part of the disk \citep{Zhu2019} :
	
	\begin{equation}
		\begin{aligned}
			\dfrac{\tau_{o}}{\tau_{p}} = \dfrac{3}{8} \dfrac{\mu}{\sqrt{1 - \mu}} \sigma^{3/2} \cos{i_{p}}
		\end{aligned}
	\end{equation}
	
	In this scenario $\sigma$ = 1, and $i_{p}$ is the angle between the orbital angular momentum vector of the companion and the rotation axis of the jet. As this is expected to be on the order of $\beta$, we take $i_{p}$ $\simeq \beta$. We can further obtain the maximum orbital radius $r_{o}$ from:
	
	\begin{equation}
		v_{o} = \frac{2\pi r_{o}}{\tau_{o}}
	\end{equation}
	
	which allows us to estimate the maximum separation \textit{a} from $r_{o} = \mu$\textit{a}. Further, from Kepler's Third Law, the total mass of a binary system is given by:
	\begin{equation}
		\frac{M_{sys}}{M_{\odot}} = \mu^{-3} \left(\frac{r_{o}}{AU}\right)^{3}\left(\frac{\tau_{o}}{yrs}\right)^{-2}
	\end{equation}
	
	Thus given estimates for $\tau_{p}$ and $M_{sys}$, we can combine these equations to obtain the orbital velocity and binary separation as a function of $\mu$. Applying our estimate for $v_{o, max}$ then gives us a maximum value of $\mu$, which we then use to estimate $\tau_{o}$ and \textit{a}.
	
	\cite{Louvet2016} estimated a mass of 1.6 \Msun\ for Th~28, using the $^{12}$CO peak velocity separations and spatial extent of the Th~28 disk. We note that \cite{Comeron2010} derived a central mass of 0.6-0.9~{\Msun\ based on the maximum velocity of H$\alpha$ detected from infalling gas in the accretion columns of the jet source. However, we take the \cite{Louvet2016} finding as accurate for an estimate of the total central mass.

    Using our estimated $v_{j}$ = 270 \kms\ and 365 \kms\ for the red- and blue-shifted jets, the mean precession parameters from the red-shifted jet give $v_{o, max}$ = 3.0 \kms\ and $\mu_{max}$ = 0.041. We find $M_{c}$ = 70 $M_{Jup}$ with an orbital period of 45 days and a binary separation of 0.3 au. 
    
    While these are not unreasonable orbital parameters for a sub-stellar companion \citep{Almeida2017, Garufi2019}, only $\sim$0.1 $\%$ of solar-type stars are thought to possess a brown dwarf companion within 3 au (the `brown dwarf desert', \cite{Luhman2007}). However, we should also note that our estimated parameters are upper limits limits on both the mass and binary separation, based on our upper bounds for $v_{o}$. The true orbital velocity is expected to be significantly smaller. Recent high-resolution observations of DO Tau also show a similarly small-scale wiggle which may be the signature of a planetary-mass companion in a very close ($<$ 1 au) orbit \citep{Erkal2021}.
    
    So how accurate are the results of the model fits? One of the most important parameters for both the fitted model and derived orbital parameters is the velocity of the jet. We derive the velocity in each lobe using the proper motions of their knots, which for the red-shifted jet is well-constrained by multiple measurements to 0\farcs34 - 0\farcs35. However, in the blue-shifted jet we have only one proper motion measurement with uncertainty 0\farcs1~yr$^{-1}$, making this the greatest source of uncertainty in the jet velocities.
    
    We therefore consider our results if the velocity of the blue-shifted jet is varied by $\pm$ 0\farcs05-0\farcs1, corresponding to $\pm$ 30-60 \kms\ (or approximately $v_{o, max}$ $\pm$ 0.1-0.2 \kms). We find that a slightly lower velocity ($v_{j, blue}$ = 330 \kms) gives similar or slightly better fits to the jet shape, while in all other cases there is a poorer fit (with average $\chi^{2}_{red}$ increasing from $\leq$ 1.1 to 1.3-1.6). However overall there is little impact on our fitted parameters. At all velocities we find $\beta$ = 0.6$^{\circ}$, with $\tau_{p}$= 7.7-8.1 years and $v_{o, max}$ = 2.8-3.0 \kms\, giving consistent companion masses of 65-75 $M_{Jup}$} and \textit{a} = 0.28-0.3 au. In no case does this change the overall findings of the models.
    
    We further consider the impact of our assumed distance of 160 pc for Th~28. Taking a 10$\%$ uncertainty on this value, we examine the results at assumed distances of 144-176 pc. This has no impact on the fitted half-opening angles or precession period, however it corresponds to a small variation in orbital velocity. At \textit{D} = 144 pc, $v_{o, max}$ = 2.7 \kms\, with a companion mass of 63 $M_{Jup}$ and \textit{a} = 0.27 au; conversely at \textit{D} = 176 pc, $v_{o, max}$ = 3.3 \kms, with companion mass 85 $M_{Jup}$ and \textit{a} = 0.34 au. Again this has a minimal impact on the overall findings.
      
    An additional concern is whether the variation of the off-axis position of the jet can be determined more accurately. While the SNR ratios of the inner portion of the red-shifted micro-jet are very high (reflected in the size of error bars in Figure \ref{fig:wigglefit}), we discuss in Section \ref{subsection:centroids} the added uncertainty at these positions due to contributions from other sources of flux than the jet. Additionally, the fitted offsets are very small relative to the seeing during the observations (0\farcs9). The off-axis distance of the jet (and hence the half-opening angle $\beta$) may therefore be underestimated. Ideally further imaging of the micro-jets with adaptive-optics corrections would allow a more accurate estimate of the wiggle parameters, as well as helping to elucidate the kinematics of the jet. 
    
    Th~28 is one of the best candidates for optical detection of a rotating stellar jet, with consistent transverse velocity gradients of approximately 10~km s$^{-1}$, detected in multiple optical emission lines by HST observations \citep{Coffey2004}. Both jets are observed to rotate in the same sense (clockwise, looking down the axis of the blue jet toward the source) which would be required for rotation as a signature of MHD jet launching. However, measurements of the Th~28 disk established that the disk rotates counter to the detected direction of the jet rotation, suggesting that this jet is not a stationary, axisymmetric flow \citep{Louvet2016} which might confound the detection of rotation stemming from jet launching \citep{Cerqueira2006}. The detection here of a wiggle in the red-shifted jet presents a possible cause of the break from axisymmetry. It is also consistent with the finding in \citet{Coffey2004} of a small ($<$ 0\farcs02) offset between the jet axis and the peak of the HVC flow when measured 0\farcs3 from the source. This offset was detected in both lobes of the jet with an equal magnitude but opposite direction from the jet axis; this is consistent both with the symmetry of the wiggle we observe, and the size of the offsets we find within 0\farcs4 of the source.

	Overall we conclude that our study of the jet axis suggests that Th~28 may possess a brown dwarf or massive planetary companion in a very close orbit. Another possibility is that this precession is not caused by the presence of an orbiting companion. Some models of jet-launching disks suggest that the interaction of the magnetic fields in the disk may produce a warped, precessing inner disk with a similar effect on the jet to precession induced by a companion \citep{Lai2003}.

    \section{Conclusions}

    We have presented MUSE observations of the Th~28 micro-jets and investigated their morphological and kinematic structure in selected emission lines. Our main findings are summarized as follows:

\begin{enumerate}

\item{The detection of [O III]$\lambda$5007 emission from the blue-shifted micro-jet of Th~28, especially the peak of the bow shock M2 at -8\farcs8 from the source, points to this jet containing strong shocks with $v_{shock}$ $>$ 100 \kms, in contrast to the red-shifted micro-jet. [O III]$\lambda$5007 emission is also seen in the central peak of the HHW bow shock, which will be discussed in a future paper.}

\item{ We identify six new knots within +/- 20 \arcsec\ of the source as well as the previously reported knot HHW$_{2}$. We find proper motions of the innermost knots as well as HHW$_{2}$ to be 0\farcs47 $^{-1}$ in the blue-shifted knot and 0\farcs34-0.35 yr$^{-1}$ in the red-shifted knots. The knots on either side show a roughly symmetrical arrangement with similar spacing either side of the star, suggesting that they are ejected at similar times from the source on a 10-15 year timescale.}

\item{The micro-jets close to the source show a wiggle indicating precession with a clear point-symmetry between the jet lobes. We find consistent fits to the precession parameters on both sides, with a period of 7.9 years and a half-opening angle $\beta$ of 0.6$^{\circ}$. We find that the shape of the precessing jet is compatible with the presence of a brown dwarf companion $\leq$ 70 $M_{Jup}$, with $\tau_{o} \leq$ 50 days, at a binary separation $\sim$ 0.3 au.}

\end{enumerate}

    These results demonstrate the active nature of the Th~28 bipolar jet, with a complex structure evolving on timescales of a few years. The observed precession may offer an alternative cause for the rotation signatures previously detected across the jet base, but the source of this small-scale precession remains unclear. Th~28 is thus an attractive target for future observations to better elucidate the processes that shape its jets.

\begin{acknowledgements} 
AM and EW wish to acknowledge the support of the Irish Research Council under the Ulysses Programme. TPR would like to acknowledge funding through the ERC Advanced Grant No. 743029 Ejection Accretion Structures in YSOs (EASY).
\end{acknowledgements}

	{}
	
	\begin{appendix}
	\section{Additional Figures and Spectro-image Catalogue}
		\begin{table}
			\centering

			\caption{All the emission lines in which the jet is detected.}
			\begin{tabular}{{p{0.08\textwidth}<{\raggedright} p{0.06\textwidth}<{\raggedright} p{0.05\textwidth}<{\raggedright} p{0.05\textwidth}<{\raggedright} p{0.05\textwidth}<{\raggedright\arraybackslash}}}
				\hline \hline
				\makecell{$\lambda_{air}$ \\(\AA)}  &Species  &Blue &Red  &Bow  
				\\  
				\hline	
				4861.3	&	H I	        &Y		&Y		&Y			
				\\																				
				4958.9	&	[O III]	    &Y		&N		&Y	
				\\					
				5006.8  &	[O III]	    &Y		&N	    &Y		
				\\	
				5158.8	&	[Fe II]	    &N		&Y	    &Y	
				\\
				5197.9  &   [NI]        &N		&Y	    &Y
				\\
				5261.6	&	[Fe II]	    &N	    &Y	    &Y		
				\\
				5754.6	&	[N II]	    &N		&Y	    &N	
				\\																				
				5875.9	&	He I	    &N		&Y	    &N		 
				\\																				
				6300.3	&	[O I]	    &Y		&Y	 	&Y	
				\\																				
				6363.7	&	[O I]	    &Y		&Y		&Y		
				\\											
				6548.9	&	[N II]	    &Y		&Y		&Y			
				\\																				
				6562.8	&	H I			&Y  	&Y      &Y			
				\\																				
				6583.5	&	[N II]	    &Y		&Y		&Y			
				\\
				6678.2	&	He I	    &N		&Y		&N			
				\\																				
				6716.4	&	[S II]	    &Y		&Y		&Y			
				\\																				
				6730.8	&	[S II]	    &Y		&Y		&Y			
				\\
				7065.2	&	He I	    &N		&Y		&N			
				\\
				7135.8	&	[Ar III]	&Y		&Y		&N			
				\\																				
				7155.2	&	[Fe II]	    &N		&Y		&N		
				\\
				7172	&	[Fe II]	    &N		&Y		&N			
				\\																		
				7291.5	&	[Ca II]	    &N		&Y		&Y	
				\\
				7319	&	[O II]	    &Y		&Y		&Y			
				\\																		
				7323.8	&	[Ca II]   	&N		&Y		&Y			
				\\
				7330.2	&	[O II]	    &Y		&Y		&Y			
				\\																		
				7377.8	&	[Ni II]	    &N		&Y		&Y	  
				\\																		
				7452.5	&	[Fe II]	    &N	    &Y	 	&N	   
				\\
				7636.2	&	[Fe II]	    &N		&Y		&N			
				\\
				8127.1	&	[Cr II]	    &N		&Y		&N			
				\\
				8498	&	Ca II	    &N		&Y		&Y		
				\\
				8617	&	[Fe II]	    &N		&Y		&Y				
				\\																				
				8891.9	&	[Fe II]	    &N		&Y		&Y			
				\\																				
				9051.9	&	[Fe II]	    &N		&Y		&N		
				\\
				9226.6	&	[Fe II]	    &N		&Y		&N	
				\\																				
				9267.6	&	[Fe II]	    &N		&Y		&N
				\\
				\hline
			\end{tabular}
			\label{em_lines}
		\end{table}

		\begin{figure*}
			\centering
			\includegraphics[width=18cm, trim= 1cm 0cm 0cm 0cm, clip=true]{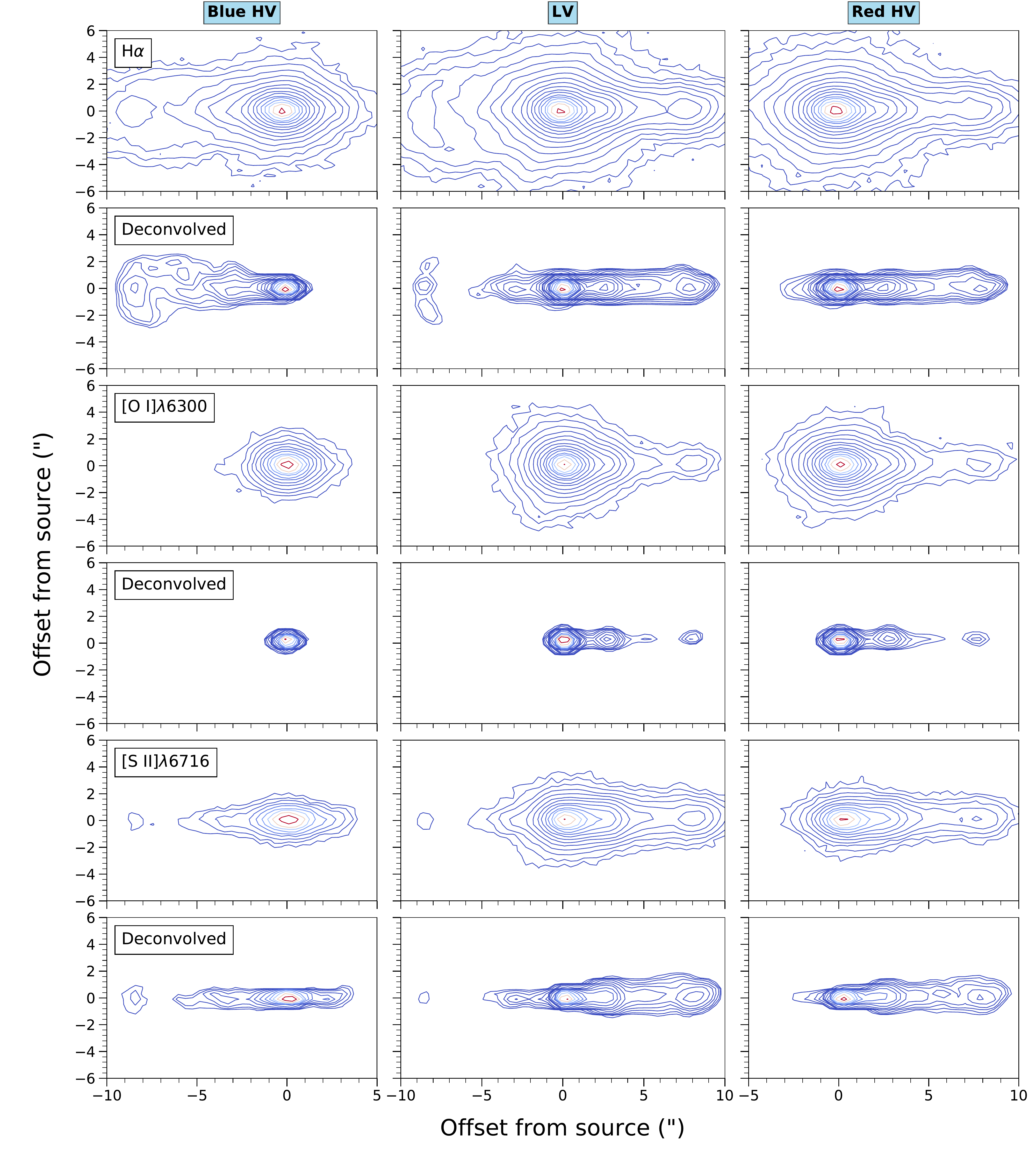}
			\caption{Spectroimages showing selected emission lines before and after Lucy-Richardson deconvolution (20 iterations), shown in three channels: the blue-shifted HVC (-150 to - 90 \kms); centre, the low-velocity channel (+/- 90 \kms); and the red-shifted HVC (+90 to -150 \kms). The inner jet region in each emission line is shown first with the same region after deconvolution shown below it. Contours begin at 3-$\sigma$ of the background noise in the unconvolved images, and increase as a factor of $\sqrt{3}$.}
			\label{fig:decon_cat1}     
		\end{figure*}

	\begin{figure*}
		\centering
		\includegraphics[width=17cm, trim= 1.2cm 1cm 0cm 0cm, clip=true]{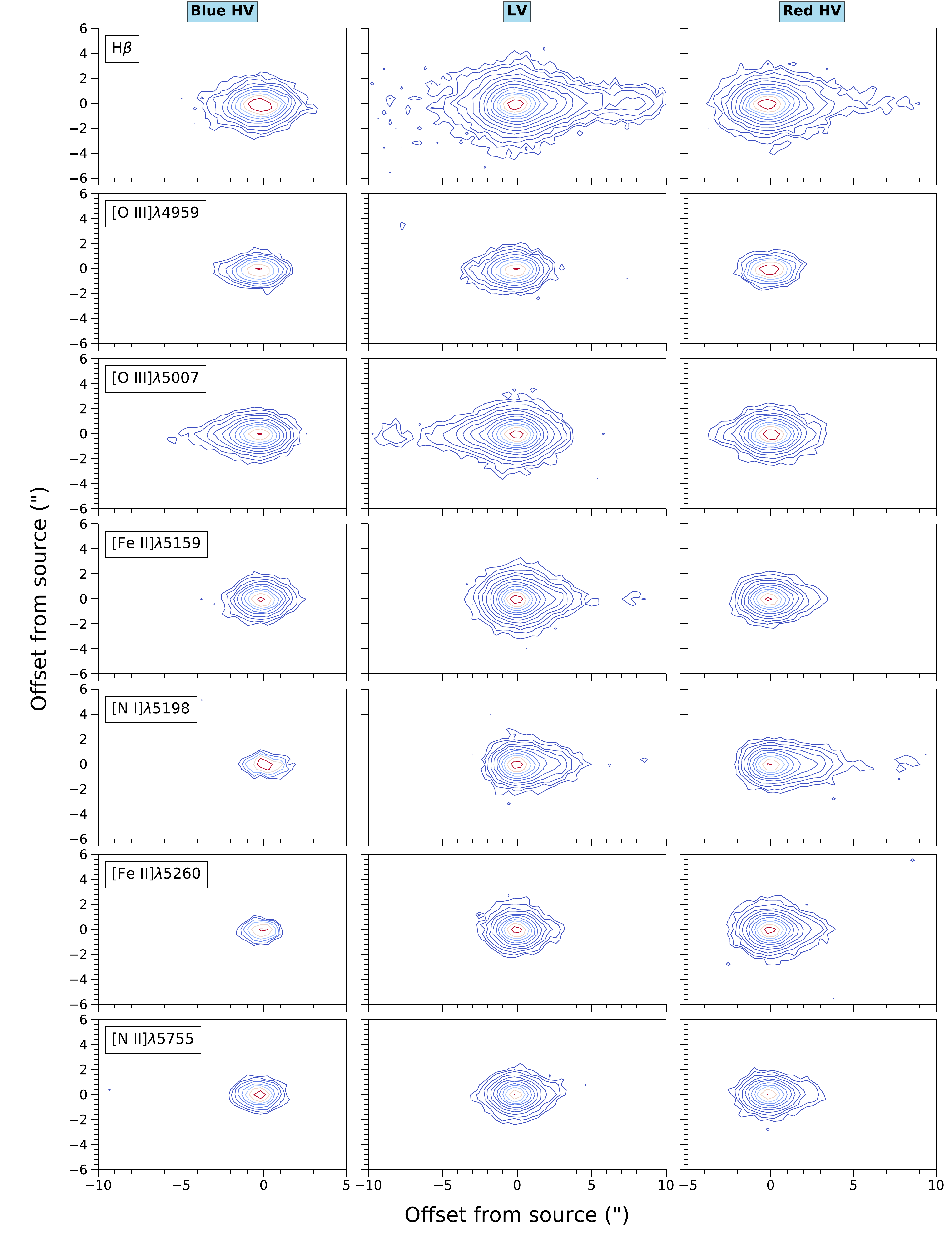}
		\caption{Spectroimages showing selected emission lines shown in three channels: the blue-shifted HV (-150 to - 90 \kms); centre, the low-velocity channel (+/- 90 \kms); and the red-shifted HV (+90 to -150 \kms). Contours begin at 3-$\sigma$ of the background noise and increase as factors of $\sqrt{3}$.}
		\label{fig:spec_cat1}     
	\end{figure*}

	\begin{figure*}
		\centering
		\includegraphics[width=17cm, trim= 1.2cm 1cm 0cm 0cm, clip=true]{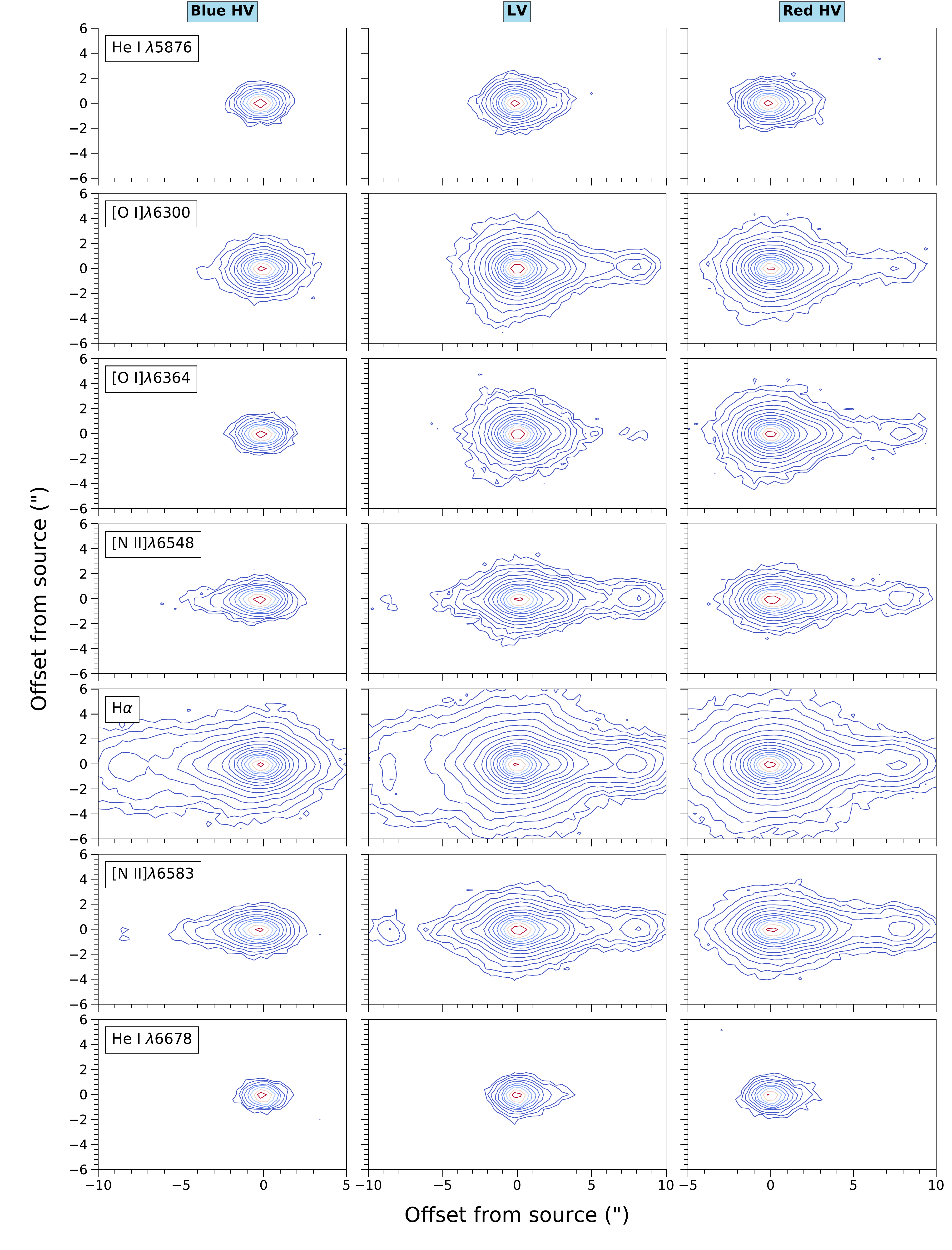}
		\caption{Spectroimages showing selected emission lines shown in three channels: the blue-shifted HV (-150 to - 90 \kms); centre, the low-velocity channel (+/- 90 \kms); and the red-shifted HV (+90 to -150 \kms). Contours begin at 3-$\sigma$ of the background noise and increase as factors of $\sqrt{3}$.}
		\label{fig:spec_cat2}     
	\end{figure*}

	\begin{figure*}
		\centering
		\includegraphics[width=17cm, trim= 1.2cm 1cm 0cm 0cm, clip=true]{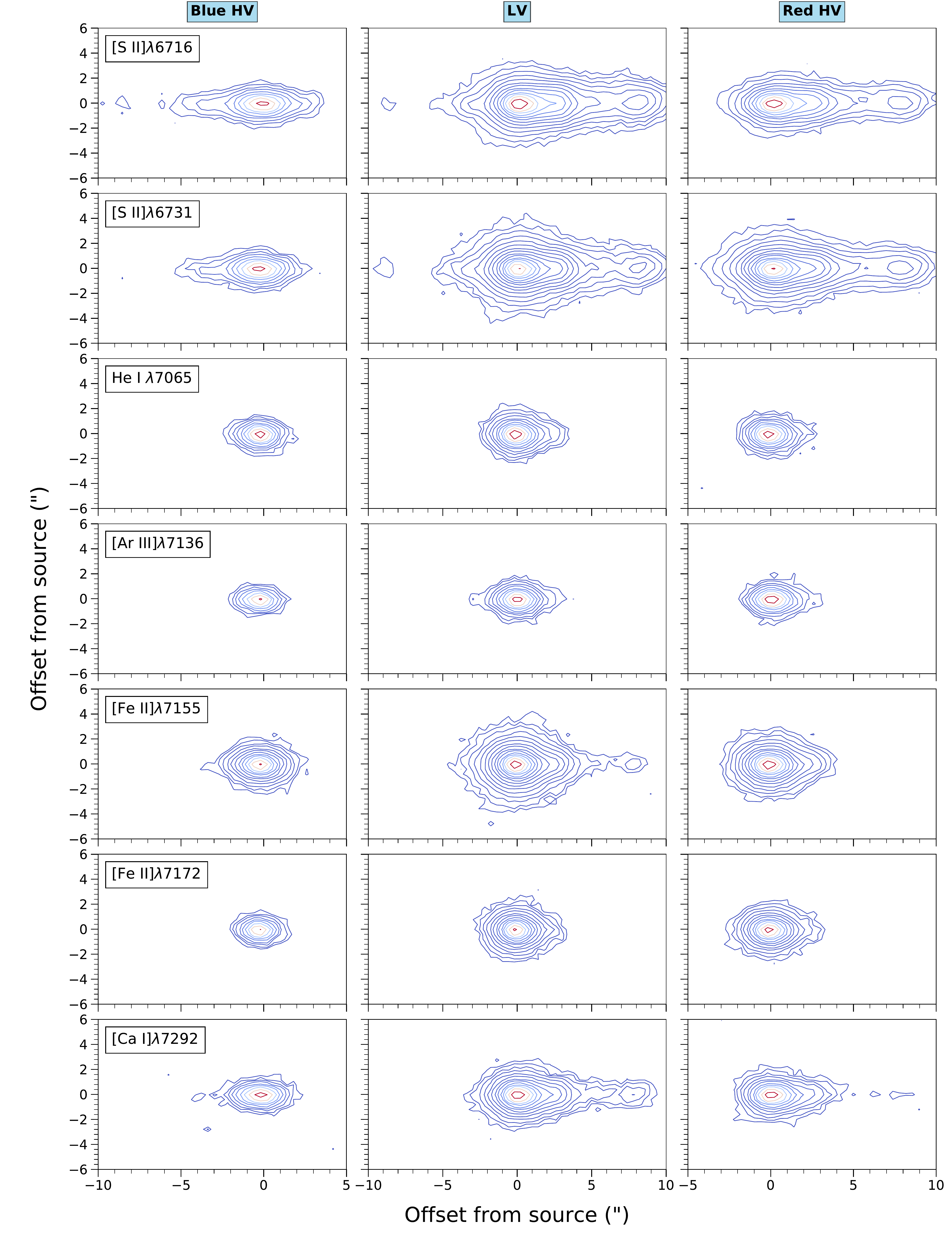}
		\caption{Spectroimages showing selected emission lines shown in three channels: the blue-shifted HV (-150 to - 90 \kms); centre, the low-velocity channel (+/- 90 \kms); and the red-shifted HV (+90 to -150 \kms).  Contours begin at 3-$\sigma$ of the background noise and increase as factors of $\sqrt{3}$.}
		\label{fig:spec_cat3}     
	\end{figure*}

	\begin{figure*}
		\centering
		\includegraphics[width=17cm, trim= 1.2cm 1cm 0cm 0cm, clip=true]{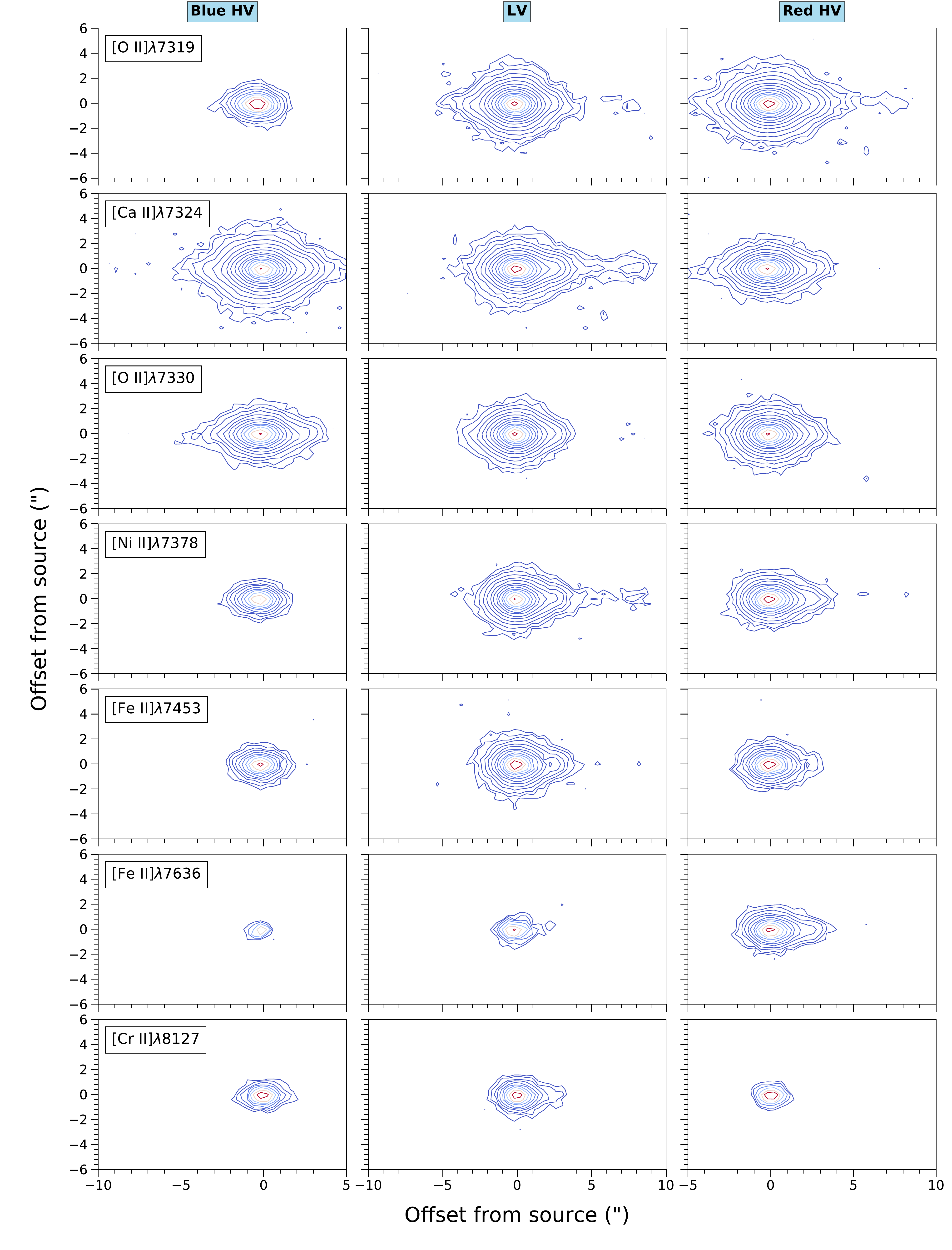}
		\caption{Spectroimages showing selected emission lines shown in three channels: the blue-shifted HV (-150 to - 90 \kms); centre, the low-velocity channel (+/- 90 \kms); and the red-shifted HV (+90 to -150 \kms).  Contours begin at 3-$\sigma$ of the background noise and increase as factors of $\sqrt{3}$.}
		\label{fig:spec_cat4}     
	\end{figure*}

	\begin{figure*}
		\centering
		\includegraphics[width=17cm, trim= 1.2cm 1cm 0cm 0cm, clip=true]{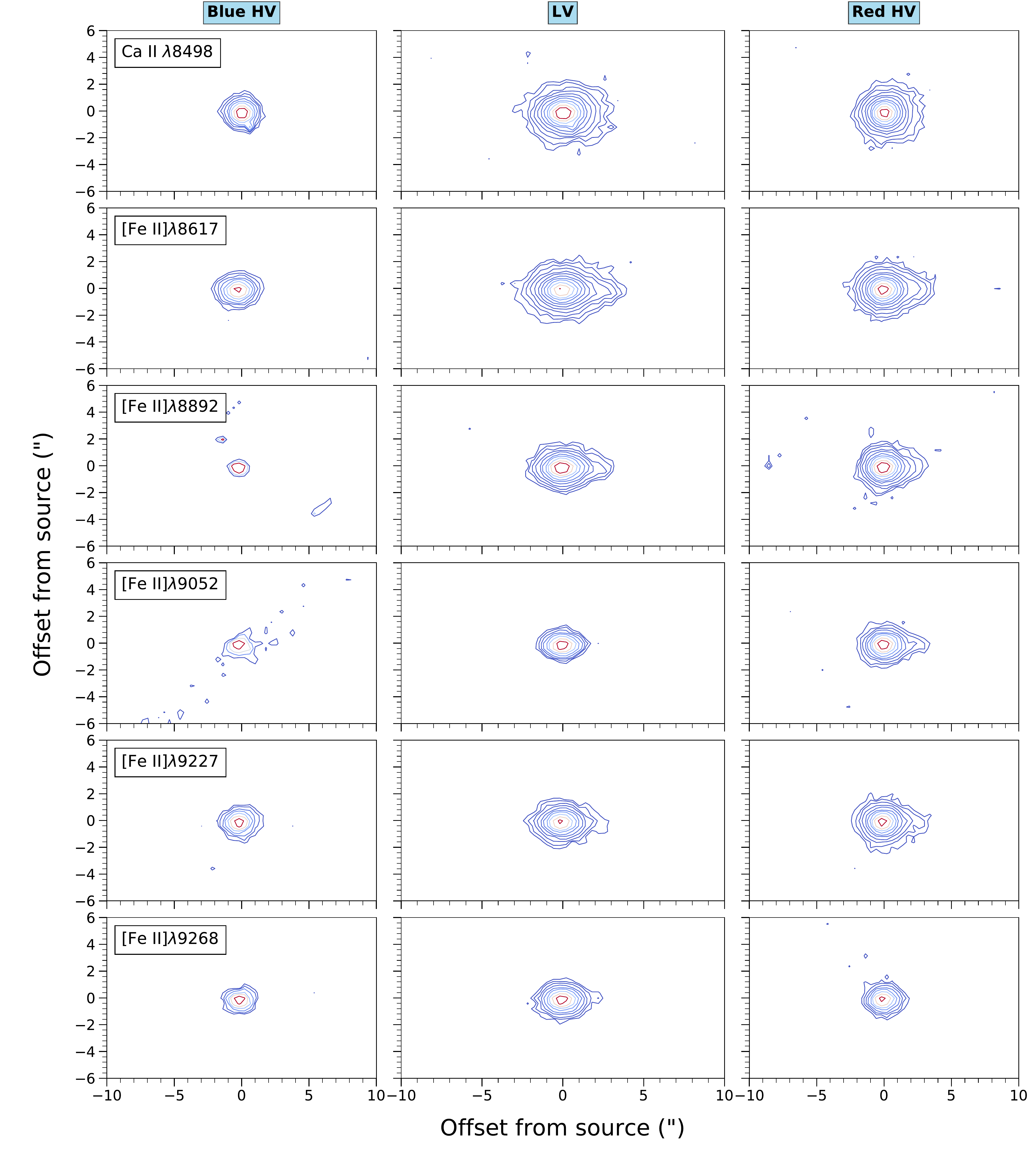}
		\caption{Spectroimages showing selected emission lines shown in three channels: the blue-shifted HV (-150 to - 90 \kms); centre, the low-velocity channel (+/- 90 \kms); and the red-shifted HV (+90 to -150 \kms). Contours begin at 3-$\sigma$ of the background noise and increase as factors of $\sqrt{3}$.}
		\label{fig:spec_cat5}     
	\end{figure*}

	\end{appendix}

\end{document}